\documentclass[twocolumn]{aastex631}

\usepackage{longtable}
\usepackage{amsmath}
\usepackage{xcolor}
\usepackage{natbib}
\usepackage{CJK}

\defcitealias{Eadie2022}{Paper~I}

\submitjournal{ApJ}

\shorttitle{The HERBAL model}
\shortauthors{Berek et al.}

\begin{document}

\title{The HERBAL model: A hierarchical errors-in-variables Bayesian lognormal hurdle model for galactic globular cluster populations}

\correspondingauthor{Samantha Berek}
\email{sam.berek@mail.utoronto.ca}

\author[0000-0001-7549-5560]{Samantha C. Berek}
\affil{David A. Dunlap Department of Astronomy \& Astrophysics, University of Toronto, 50 St George Street, Toronto, ON M5S 3H4, Canada}

\author[0000-0003-3734-8177]{Gwendolyn M. Eadie}
\affiliation{David A. Dunlap Department of Astronomy \& Astrophysics, University of Toronto, 50 St George Street, Toronto, ON M5S 3H4, Canada}
\affiliation{Department of Statistical Sciences, University of Toronto, 9th Floor, Ontario Power Building, 700 University Ave, Toronto, ON M5G 1Z5, Canada}

\author[0000-0003-2573-9832]{Joshua S. Speagle (\begin{CJK*}{UTF8}{gbsn}沈佳士\ignorespacesafterend\end{CJK*})}
\affiliation{David A. Dunlap Department of Astronomy \& Astrophysics, University of Toronto, 50 St George Street, Toronto, ON M5S 3H4, Canada}
\affiliation{Department of Statistical Sciences, University of Toronto, 9th Floor, Ontario Power Building, 700 University Ave, Toronto, ON M5G 1Z5, Canada}
\affiliation{Dunlap Institute for Astronomy \& Astrophysics, University of Toronto, 50 St George Street, Toronto, ON M5S 3H4, Canada}
\affiliation{Data Sciences Institute, University of Toronto, 17th Floor, Ontario Power Building, 700 University Ave, Toronto, ON M5G 1Z5, Canada}

\author[0000-0001-8762-5772]{William E. Harris}
\affiliation{Department of Physics \& Astronomy, McMaster University, 1280 Main Street West, Hamilton, L8S 4M1, Canada}

\begin{abstract}
Galaxy stellar mass is known to be monotonically related to the size of the galaxy's globular cluster (GC) population for Milky Way sized and larger galaxies. However, the relation becomes ambiguous for dwarf galaxies, where there is some evidence for a downturn in GC population size at low galaxy masses. Smaller dwarfs are increasingly likely to have no GCs, and these zeros cannot be easily incorporated into linear models. We introduce the Hierarchical ERrors-in-variables Bayesian lognormAL hurdle (HERBAL) model to represent the relationship between dwarf galaxies and their GC populations, and apply it to the sample of Local Group galaxies where the luminosity range coverage is maximal. This bimodal model accurately represents the two populations of dwarf galaxies: those that have GCs and those that do not. Our model thoroughly accounts for all uncertainties, including measurement uncertainty, uncertainty in luminosity to stellar mass conversions, and intrinsic scatter. The hierarchical nature of our Bayesian model also allows us to estimate galaxy masses and individual mass-to-light ratios from luminosity data within the model. We find that 50\% of galaxies are expected to host globular cluster populations at a stellar mass of $\log_{10}(M_*)=6.996$, and that the expected mass of GC populations remains linear down to the smallest galaxies. Our hierarchical model recovers an accurate estimate of the Milky Way stellar mass. Under our assumed error model, we find a non-zero intrinsic scatter of $0.59_{-0.21}^{+0.3}$ (95\% credible interval) that should be accounted for in future models.


\end{abstract}

\section{Introduction} \label{sec:intro}

Globular clusters (GCs) are old and massive star clusters which are found in almost all galaxies, excluding the smallest dwarfs \citep[see][for a review]{Harris2010, Beasley2020}. For Milky Way sized and larger galaxies, an empirical linear relationship exists between the halo mass of a galaxy ($M_{halo}$) and the mass of its GC cluster system ($M_{gcs}$)  \citep{Blakeslee1997, Harris2013, Hudson2014, Forbes2018}. The observed relationship between galactic stellar mass ($M_*$) and $M_{gcs}$ has more scatter than that with $M_{halo}$ and is nonlinear when considered over the entire range of galaxy luminosities \citep{Harris2013,deSouza2015}, even though GCs are stellar systems and we would expect their formation to be strongly tied to a galaxy's baryonic mass properties. 

Multiple studies have used simulations and theoretical modeling to investigate the $M_{halo}$ - $M_{gcs}$ relation and its origin \citep{ElBadry2019, Choksi2019, Bastian2020, Chen2023}. However, observational measurements of galaxy halo masses are limited, while stellar masses are more easily obtained through luminosity measurements and conversions with a mass-to-light ratio (M/L). Therefore, a better understanding of the shape of the $M_*$ - $M_{gcs}$ relationship is important in furthering our knowledge of galaxy and GC formation and co-evolution in the observed universe. 

Neither the $M_{halo}-M_{gcs}$ nor the $M_*-M_{gcs}$ relation are well understood for dwarf galaxies. A downturn in the otherwise 1:1 $M_{halo}$ - $M_{gcs}$ relation has been observed in some observational and modeling studies \citep[e.g.][]{Choksi2019, Bastian2020}, but others have concluded that the relationship is linear down to low galaxy masses \citep{Spitler2009, Hudson2014, Harris2017}. It is unclear whether the downturn seen in some studies is a real change in globular cluster system properties, due to an intrinsic slope change in the stellar mass halo mass relation, or is a product of the inclusion or exclusion of galaxies without globular cluster populations. To fully understand how the $M_*$ - $M_{gcs}$ relation behaves in the low mass regime, a thorough investigation of the transition mass region, where galaxies shift from not having GC populations to having them, is needed. 

\citet{Eadie2022} (hereafter \citetalias{Eadie2022}) undertook the first comprehensive analysis of the transition mass region of dwarf galaxies. It tested various empirical models of the GC populations of dwarf galaxies, and concluded that a Bayesian lognormal hurdle model was the best description of the data. This class of model is a combination of a logistic (probabilistic) model and a linear model, allowing for the simultaneous estimation of a binary cluster label (whether a galaxy does or does not have globular clusters) and an estimation of the expected GC population mass. This exploratory result suggests that the hurdle model should be investigated more thoroughly.  

In this follow-up paper to \citetalias{Eadie2022}, we expand the Bayesian lognormal hurdle model into an errors-in-variables model that has the ability to account for observational uncertainties as well as fully describe the uncertainties in both observed luminosities and the calculated mass-to-light (M/L) ratios necessarily for estimating stellar masses from luminosity data. First, in Section \ref{sec:definitions}, we discuss our criteria in choosing galaxies and globular clusters. Section \ref{sec:sample} describes the observational sample used for this study. A comprehensive sample of Local Group galaxies and their globular clusters is compiled and available for download. In Section \ref{sec:hurdle}, we describe the Bayesian lognormal hurdle model for galaxy luminosities and expand it into an errors-in-variables model that incorporates observational uncertainties. In Section \ref{sec:lumtomass}, we develop a hierarchical model (the HERBAL model) that estimates galaxy masses and accounts for multiple sources of uncertainty. Section \ref{sec:disc} discusses sources of scatter in the model, the ability of the hierarchical model to learn about M/L ratios as well as galaxy masses, its robustness to outliers, and the model-derived properties of the Milky Way. Section \ref{sec:conc} summarizes our conclusions and future work, including the potential of incorporating other galaxy parameters, mixture components, and error distributions. 

\section{Defining the sample} \label{sec:definitions}

Neither galaxies nor globular clusters are distinct sets of objects that can be defined by the observational data that we possess. Small dwarf galaxies and large star clusters overlap in many of their physical properties, as do globular clusters and other types of star clusters. To compile a complete census of galaxies and their globular clusters, therefore, we first must define both of these classes of objects. 

\subsection{What defines a galaxy?} \label{subsec:galaxydef}

The large galaxies in the Local Group are unambiguous, but many of the small and faint dwarf galaxies are demarcated in surveys as galaxy candidates. These smallest objects sit in the overlap mass region between star clusters and galaxies. The distinction between a galaxy and a GC is based on dark matter - galaxies are dark matter sub-halos, while GCs are simply groups of stars, with no accompanying dark matter overdesnsity.

There are two main ways that objects small enough to plausibly be star clusters can be classified as a galaxy. The first is with spectroscopic velocity dispersion data: a large dispersion paired with a faint luminosity indicates a very high M/L ratio and an over-density of dark matter, which classifies the object as a galaxy. The second relies on metallicity data and an assumption: in a small galaxy, a dark matter overdensity is necessary to create a deep enough potential well that supernovae ejecta become graviationally bound to the system, getting recycled into future stars instead of being lost to intergalactic space. In other words, high metallicity stars are assumed to only exist in systems with a deep potential well and dark matter, classifying systems with these stars as galaxies instead of clusters. 

Spectroscopic velocity dispersion data directly points to the presence of a dark matter overdensity in small galaxies, while high metalliticy stars indirectly points to a deep potential well and therefore dark matter. We choose to classify galaxies conservatively, and define galaxies as objects in one of the following two categories:
\begin{itemize}
    \item Any object large enough that it could not be a star cluster, and that is called a galaxy in multiple Local Group surveys \citep[i.e.][]{McConnachie2012, Simon2019, DrlicaWagner2020}.
    \item An ultra-faint dwarf with a spectroscopic velocity dispersion measurement that implies a M/L ratio far larger than would be expected based on just a stellar component alone.
\end{itemize}

\subsection{What defines a globular cluster?} \label{subsec:gcdef}

Globular clusters are distinguished from other star clusters by a variety of features, including their age, metallicity, and location in their host galaxy (see \citet{Harris2010} and references therein for a review). In the Milky Way, which is the best-studied globular cluster system, there is a relatively clear delineation between open clusters and globular clusters. Open clusters are young, relatively small, loosely gravitationally bound, metal rich, and found in the galactic disk. Globular clusters are in many ways the opposite: they are old, metal poor, relatively massive, tightly gravitationally bound, and mostly found in the halo. 

In other galaxies, however, the separation between these two populations of star clusters becomes less clear. In small dwarf irregular galaxies, for example, there is no disk and halo to make a distinction in terms of location. In galaxies with more continuous star formation histories than the Milky Way, clusters have been formed over a wide range of ages, which also gives them a more continuous range of metallicities. In M33, for example, star clusters exist along a continuum of ages from under 1 Gyr to 13 Gyr \citep{Fan2014}.

For consistency, we use data from the Milky Way to define an age cut for globular clusters, which we use as our selection criteria.  We choose a value based on the results of \citet{VandenBerg2013} and \citet{Leaman2013}, which found that all the MW clusters in their sample are, within error bars, older than 8 Gyr. Therefore, our globular cluster definition is: 
\begin{itemize}
    \item Any star cluster older than 8 Gyr.
\end{itemize}

\section{Observational Sample} \label{sec:sample}

\begin{figure*} 
    \centering
    \includegraphics[width=0.95\textwidth]{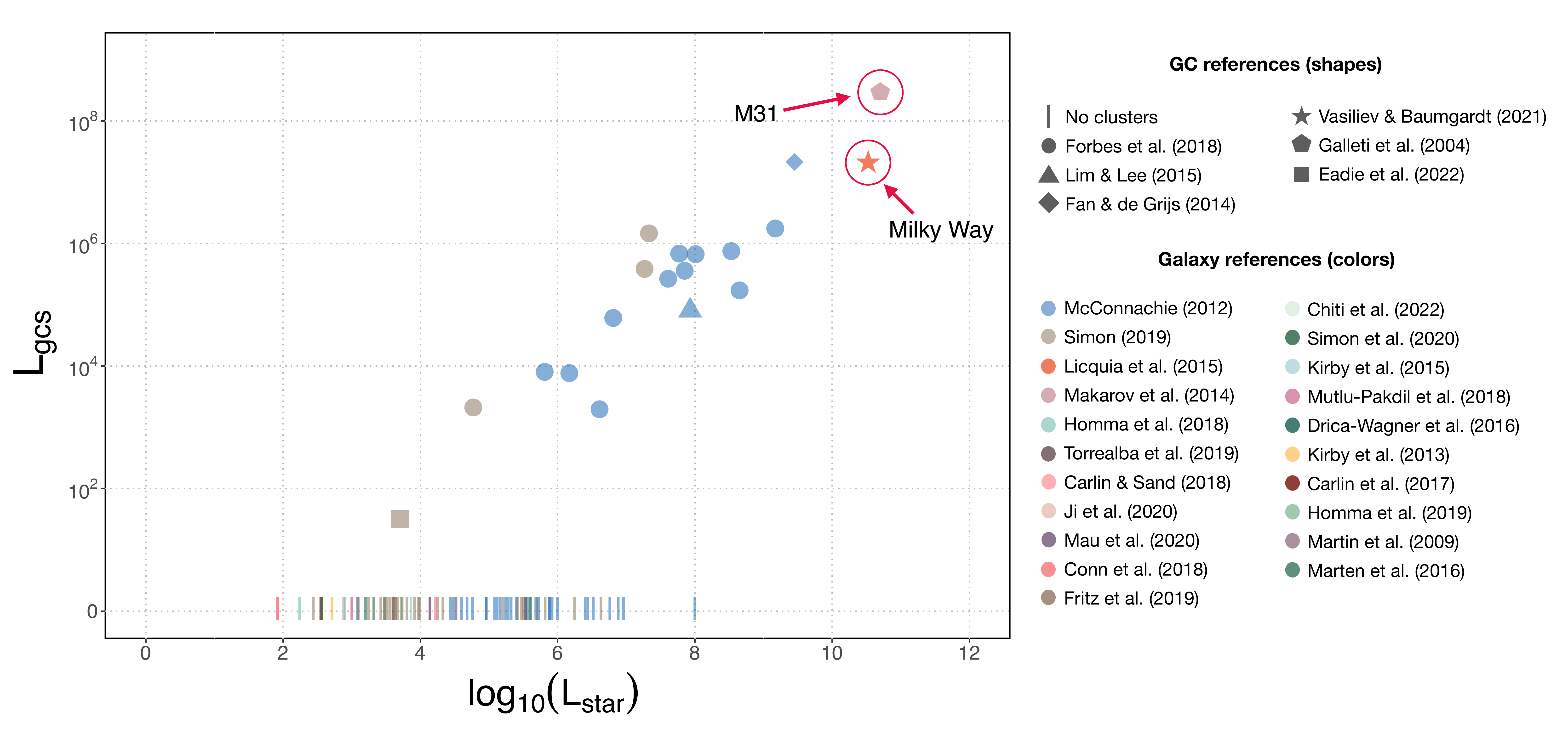}
    \caption{Galaxy luminosities and globular cluster system luminosities of our full Local Group data sample, including galaxy candidates. The shape of points corresponds to the source of their globular cluster luminosity data (galaxies without GCs are shown as tick marks). Points are colored based on the source of the galaxy luminosity data. Note that a subset of galaxies have multiple references for galaxy data; some of the dwarf galaxies have information from \citet{Simon2019} as well as another source. These points are colored according to the non-Simon source.}
    \label{fig:data_references}
\end{figure*}

A complete census of galaxies and their cluster populations, which is needed to avoid bias in a model fit, is only possible in the nearby universe. This is due to the relatively low luminosities of globular clusters and frequent contamination in galaxies distant enough that GCs appear as unresolved, point-like sources. Therefore, we use the Local Group members for the purposes of this initial study. The Local Group has been studied extensively and GC populations are resolved and almost complete. 

The sample is compiled from a variety of catalogs and sources of both Local Group galaxies and globular clusters. We use the definitions of both galaxies and GCs described in Section \ref{sec:definitions} in making the sample. Our data, which consists of luminosities of both galaxies and their cluster populations, uncertainties, and GC counts, is previewed in Table \ref{tab:sample_preview}. The full table is available in Appendix \ref{sec:app1}. The compilation of data is illustrated in Figure \ref{fig:data_references}. Points are colored according to the source that the galaxy data was taken from, and shapes correspond to GC data sources. 

\subsection{The Local Group Galaxies} \label{subsec:sample_galaxies}

We compile a preliminary list of Local Group members from \citet{Simon2019}, \citet{DrlicaWagner2020} and \citet{McConnachie2012}, and we pull $M_V$ absolute magnitude data and uncertainties for the galaxies from these sources. The absolute magnitudes are then converted to $\log$\footnote{Throughout, all $\log$ quantities are $\log_{10}.$} luminosities using the solar absolute magnitude $M_{\odot,V}=4.83$ and the equation
\begin{equation}
    \log{L_*} = \frac{|M_{\odot,V}-M_{*,V}|}{2.5}
\end{equation}
for a galaxy with an absolute magnitude $M_{*,V}$. We divide uncertainties in $M_{*,V}$ by $2.5$ to obtain uncertainties in $\log{L_*}$. 

Five galaxies in the sample have different upper and lower uncertainty values in the literature. We standardize these uncertainties by adopting the larger value as a symmetric uncertainty. This is done so that errors can be treated as symmetric distributions.

We conduct a literature search of objects listed as galaxy candidates in the aforementioned surveys. Objects for which spectroscopic velocity dispersions have been measured and match those expected of a dark matter over-density are moved to the sample of confirmed galaxies. For the objects that do not yet have clear confirmation, we make a secondary list of galaxy candidates and run our model both with and without these candidates. The confirmed sample contains a total of 81 galaxies, while the secondary candidate sample contains a further 17 galaxies. 

\subsection{Globular Cluster Sample} \label{subsec:sample_clusters}

Eighteen galaxies in the Local Group are known to host globular clusters. The GC populations of these galaxies are uncertain due to the same issues presented in Section \ref{sec:definitions}: there is overlap in luminosity, size, and mass between the smallest galaxies and globular clusters, making it unclear if some objects are galaxies or clusters. Additionally, although the Local Group sample is the most complete sample available due to its proximity, some faint clusters may still be undiscovered. Further, satellite galaxies interact with their companions, making it unclear if nearby clusters belong to the dwarfs or the central galaxy.

Acknowledging these constraints, we compile data on the cluster populations of the Local Group galaxies. The Milky Way cluster data is taken from the Baumgardt et al. globular cluster database, which consists of 162 galactic globular clusters with parameters derived using Gaia EDR3 proper motions, HST based stellar mass functions, and ground-based radial velocities \citep{Baumgardt2017, Sollima2017, Baumgardt2018, Vasiliev2021}. Cluster masses are estimated by comparing velocity dispersions and surface brightness profiles to a grid of N-body simulations of star clusters with varying initial conditions which are then evolved forward under conditions of stellar evolution and two-body relaxation \citep{Baumgardt2017}. We use these masses to calculate the total GC system mass and associated uncertainty for the Milky Way.

We use the Galleti cluster catalog for M31 \citep{Galleti2004}. This catalog consists of all confirmed and candidate clusters; we take all clusters that are labelled as confirmed. The catalog lists observational properties of the clusters, and so we use their apparent V-band magnitudes and reddening to compute the total mass of the GC system. We first calculate the intrinsic V-band magnitude $V_0 = V_{obs} - 3.1E(B-V)$, and then use a distance modulus of $\mu_0=24.407$ \citep{Li2021} to calculate absolute magnitude $M_V$ and log luminosity $\log{L_{\textrm{gcs}}}$. We do not have distances to individual clusters, and so we use the M31 distance for all clusters. Since clusters should be distributed fairly symmetrically about the central galaxy, we anticipate that we have overestimated and underestimated distances approximately equally. 

\vspace*{-\baselineskip}
\vspace*{-\baselineskip}
\begin{deluxetable*}{cccccccc} \label{tab:sample_preview} 
    \tablecaption{Table of Local Group Galaxies}
    \tablehead{
    \colhead{Galaxy} & \colhead{Confirmed} & \colhead{\textbf{$\log{L_*}$}} & \colhead{$e\log{L_*}$} & \colhead{$N_{GC}$} & \colhead{$M_{*,GC}$} & \colhead{Galaxy Ref.} & \colhead{GC Ref.}
    }
    \startdata
        And I & 1 & 6.612 &  0.04 & 1 & $3.68 \times 10^3$ & 0 & 1 \\
        Eridanus II & 1 & 4.772 &  0.12 & 1 & $4 \times 10^3$ & 1 & 1 \\
        Aquarius & 1 & 6.172 & 0.04 & 1 & $1.46 \times 10^4$ & 0 & 1 \\
        And XXV & 1 & 5.812 & 0.2 & 1 & $1.46 \times 10^4$ & 0 & 1 \\
        Pegasus & 1 & 6.812 & 0.08 & 1 & $1.15 \times 10^5$ & 0 & 1 \\
        SMC & 1 & 8.652 & 0.08 & 1 & $3.2 \times 10^5$ & 0 & 1 \\
        \vdots & & & & \vdots & & & \vdots \\
    \enddata
    \tablecomments{This table contains a preview of the Local Group dataset. For the full table, a description of columns, and list of references, see Appendix \ref{sec:app1}}
\end{deluxetable*}

The cluster system of M33 is described in \citet{Fan2014}. To separate globular clusters from other clusters, we adopt our age cut of 8 Gyr, considering all old clusters to be globular clusters. \citet{Fan2014} have also calculated masses for the clusters. They compare the observed SEDs of the clusters with \texttt{PARSEC} isocrones to estimate both ages and masses. We compile a GC system mass, along with an uncertainty, from the values in that work.

IC 10 also has a large population of star clusters. \citet{Lim2015} reports cluster luminosities and calculates ages and masses. Ages and masses are calculated by comparing UBVRI spectral energy distributions of the clusters, assuming a Salpeter initial mass function, with simple stellar population models. Only one cluster passes our age cut and is recorded as a GC.

We take GC system luminosities and masses of all other Local Group galaxies from \citet{Forbes2018}, which contains further information on individual galaxies and their cluster populations.

We convert between GC luminosities and masses using a set $M/L$ ratio,
\begin{equation}
   M/L = 1.88. 
\end{equation}
This is a standard choice for globular clusters \citep[e.g.][]{McLaughlin2005, Georgiev2010, Forbes2018} which typically cannot be improved upon without further information on individual clusters' star formation histories. We adopt a factor of two uncertainty in our cluster mass estimates to account for variation among M/L ratios of GC systems of different galaxies.

\section{The errors-in-variables Bayesian lognormal hurdle model} \label{sec:hurdle}

A Bayesian lognormal hurdle model is a type of generalized linear model (GLM) \citep{Nelder1972, McCullaugh1983}, a flexible class of models that relate the predictors to the response through a set of linear parameters and a link function. Unlike general linear models, which follow the form \boldmath$Y=\beta X$ \unboldmath and make the assumption that the response variable and residual uncertainties are normally distributed with constant variance, generalized linear models relax these assumptions and allow for a wider range of model types. They still require linear predictors \boldmath$\eta=\beta X$\unboldmath, but allow them to be transformed by an exponential link function of the mean $g(\boldsymbol\mu)=\boldsymbol\eta$. Thus, the overall form of a generalized linear model is 
\begin{equation} 
    Y = g^{-1}(\boldsymbol\eta)=g^{-1}(\boldsymbol\beta\mathbf{X}),
\end{equation}
where the linear predictors are transformed by the link function instead of being directly related to the response variable. GLMs have only occasionally been discussed and used in astronomy \citep[see][for an introduction of their use in the field]{deSouza2015, Elliot2015, deSouza2015B, hilbe2017, Hattab2019}, but their flexibility makes them well suited for a multitude of astronomical applications. 

\subsection{The Bayesian lognormal hurdle model}

The hurdle model used in \citetalias{Eadie2022} is a GLM; specifically, a lognormal hurdle, which is a combination of a logistic and linear regression (in log space). The logistic portion gives the probability that a point at some $x$ value will be either zero or nonzero, while the linear portion gives the expected $y$ value for that point if it is nonzero. 

The hurdle model can be written as a combination of a Bernoulli and normal distribution: 
\begin{align} \label{equ:hurdle_dist}
    I &\; \sim \; \mathrm{Bern}(p(x)), \nonumber \\
    Y|(I=0)& \; = \; 0, \\
    Y|(I=1)& \; \sim \; \mathcal{N}(\mu(x), \sigma). \nonumber 
\end{align}
The Bernoulli distribution is a single trial of a binomial distribution that returns $1$ or $0$ based on the probability of success $p$ for that trial. If the Bernoulli trial returns a $0$ (or failure), the model also returns a zero $y$ value for that point. If the trial returns a $1$ (or success), the $y$ value is sampled from a normal distribution with mean $\mu$ and the standard deviation $\sigma$. 

The functions $p(x)$ and $\mu(x)$ are chosen based on the scientific context. For this data, we use a logistic regression to model the probability of success or failure $p(x)$ and a linear regression to model the mean of the normal distribution for non-zero data $\mu(x)$: 
\begin{align}
    p(x) &\; = \; \frac{1}{1+e^{-(\beta_0+\beta_1 x)}}, \\
    \mu(x) & \; = \; \gamma_0 + \gamma_1 x. \nonumber
\end{align}
These are the expectation values of each part of the model, and they can be multiplied to give the overall expectation value of the full hurdle model. 

\citetalias{Eadie2022} uses this model for the stellar masses of galaxies and their globular cluster populations. Here, we instead use luminosities of galaxies and their combined GC populations \footnote{For GCs for which we have direct mass data (see Section \ref{subsec:sample_clusters}), we use an $M/L$ ratio of $1.88$ to convert masses to luminosities.}. Using luminosities is the most data-driven approach and removes uncertainty in the mass-to-light ratios which must be assumed for conversions to stellar mass. The expectation value for our hurdle model then becomes:
\begin{equation} \label{equ:hurdle}
E[\log{L_{GC}}]=\frac{1}{1+e^{-(\beta_0+\beta_1\log{L_*})}}(\gamma_0+\gamma_1\log{L_*}).  
\end{equation}
The model samples for four parameters: $(\beta_0, \beta_1)$, the parameters of the logistic portion of the model, and $(\gamma_0, \gamma_1)$, the linear parameters.

Just like in a linear regression, the $y$ values of the best-fit line are the expectation values of the response. However, because the expectation values of the hurdle model combine a linear model with a probabilistic model, we do not expect the data to cluster around the expectation values. Instead, the data separate into two sub-populations - one with $y$ values of zero, and the other with non-zero $y$ values. If the data were binned and thus the two populations of zero and non-zero responses were combined, the data would adhere to the shape of the expectation value curve. 

\subsection{An errors-in-variables hurdle model} \label{subsec:eiv}

The hurdle model used in \citetalias{Eadie2022} and described above does not include measurement uncertainties for either the predictor or response due to their being heteroskedastic and non-Gaussian. However, incorporating known uncertainties provides more information about the data, and so we improve upon \citetalias{Eadie2022} by developing an errors-in-variables version of the hurdle model. 

The measurement uncertainty of the response variable, in this case GC system luminosity, can be accounted for through the parameter $\sigma$ in the normal distribution shown in Equation \ref{equ:hurdle_dist}. The normal distribution describes the population of galaxies that do have GCs at a particular galaxy luminosity. Incorporating the uncertainty in combined cluster luminosity here implies that there is no overlap between the two populations  - that is, that there is zero probability that the galaxies with no known clusters actually have some or vice versa. A future version of this model might incorporate a mixture model to encompass the possibility of undiscovered clusters or mislabeled GCs that are actually other types of stellar clusters. 

Uncertainties in the predictor variable can be incorporated by treating the galaxy luminosity data as randomly sampled values drawn from a distribution instead of as true values. For the predictor $\log{L_{*,i}}$, the errors-in-variables model samples for $\log{L_{*,i}^{true}}$, the true galaxy luminosities, while considering the sample $\log{L_{*,i}}$ to be noisy measurements. The distribution of each $\log{L_{*,i}}$ is assumed to be normal with standard deviation $\sigma_{\log{L_{*,i}}}$:
\begin{equation}
    \log{L_{*,i}} \sim \mathcal{N}(\log{L_{*,i}^{true}},\sigma_{\log{L_{*,i}}}),
\end{equation} 
where $\sigma_{\log{L_{*,i}}}$ is the measurement uncertainty on galaxy luminosities $\log{L_{*,i}}$. This method adds $i$ parameters to the problem, but this does not significantly increase computation time using samplers such as \texttt{Stan} \citep{stan}, which utilize gradient based methods of sampling. 

\begin{figure*}
    \centering
    \includegraphics[width=0.95\textwidth]{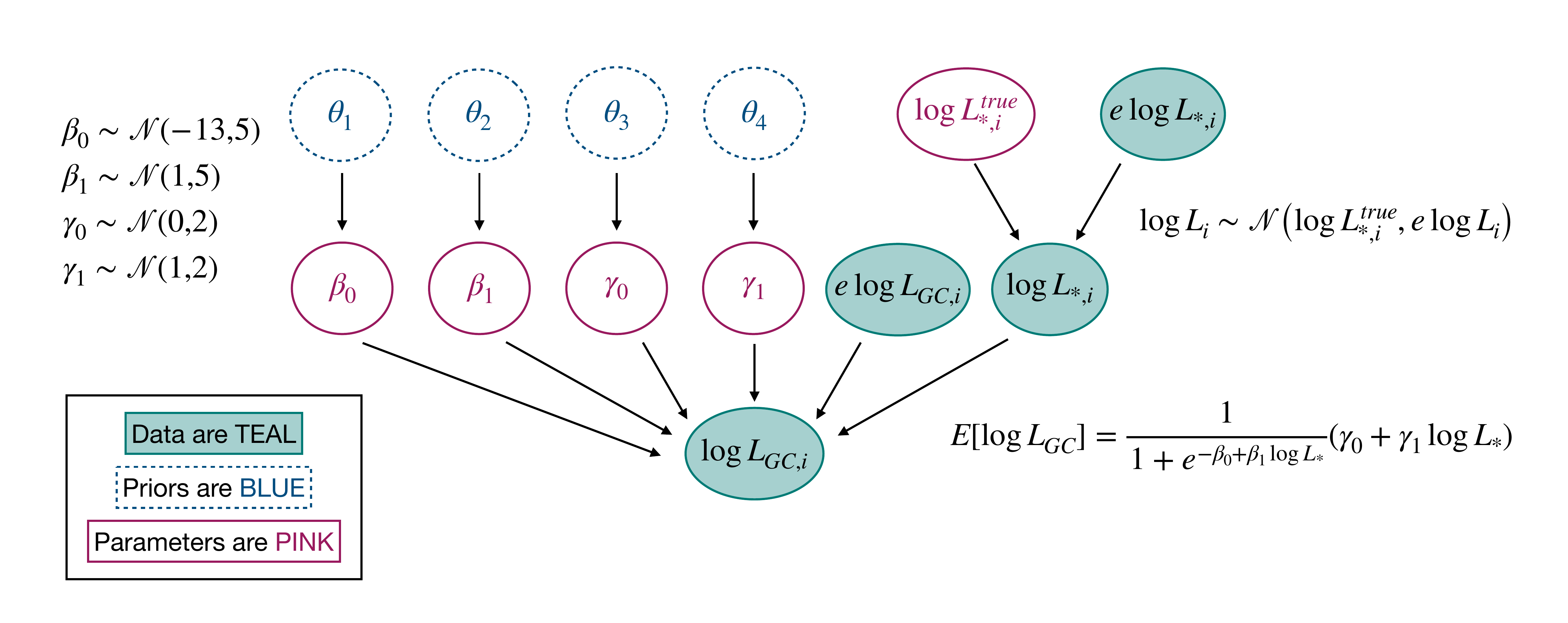}
    \caption{Directed acyclic graph (DAG) of the errors-in-variables luminosity hurdle model. Arrows follow down the steps of the model, linking inputs with responses of each distribution and leading to the response variable $\log{L_{GC,i}}$ in the bottom row. Priors for the model parameters are blue, data are teal, and model parameters are pink. The distributions that follow down the layers of the DAG are written beside each arrow.}
    \label{fig:dag_lum}
\end{figure*}

The structure of the errors-in-variables model is shown in Figure \ref{fig:dag_lum}. Figure \ref{fig:dag_lum} is a directed acyclic graph (DAG), which is a format often used in Bayesian statistics to represent hierarchical models. Arrows show which variables are necessary to calculate or estimate the others. Moving up the graph moves up levels of the model. Data are colored in teal, model parameters are in pink and priors are in blue. On the sides of the graph, the distributions and equations that link each level with the level above it are shown.

The bottom level of the DAG shows the response variable of our model, the globular cluster system luminosity $\log{L_{GC,i}}$. The next level consists of all the inputs to the hurdle model, which include the four model parameters $(\beta_0, \beta_1, \gamma_0, \gamma_1)$, the predictor variable $\log{L{*,i}}$ (galaxy luminosities), and the measurement uncertainties on the GC systems. The galaxy luminosities, in turn, have their own associated uncertainty, and so the model samples for their true values $\log{L_{*,i}^{true}}$, which adds another layer to the model. Lastly, each of the model parameters has a prior listed on the left hand side. We do not place an informative prior on the true galaxy luminosities. 

We implemented this model in \texttt{Stan} using the package \texttt{RStan} in the R Statistical Software Environment. \texttt{Stan} is a platform for statistical modeling and computation that uses the no-U turn sampler (NUTS), an optimized version of Hamiltonian Monte Carlo (HMC), which is a gradient based method of sampling \cite{hmc, nuts}. HMC allows for incredibly time-efficient sampling; our models all ran in under a minute on a standard laptop. 

\subsection{Prior sensitivity and sample selection} \label{subsec:lum_priors}

We tested a variety of priors for the four model parameters, informed by the priors used in \citetalias{Eadie2022}. Small changes in the priors did not have an overall effect on the model, and so we use the priors used in \citetalias{Eadie2022}: 
\begin{align*}
    \beta_0 &\sim \mathcal{N}(-13,5) \\
    \beta_1 &\sim \mathcal{N}(1,5) \\
    \gamma_0 &\sim \mathcal{N}(0,2) \\
    \gamma_1 &\sim \mathcal{N}(1,2) \\
\end{align*}
These priors are only weakly informative, but without much prior knowledge we prefer to leave them relatively broad. These four model (or population-level) parameters are estimated along with true values for each galaxy luminosity $\log{L_{*,i}^{true}}$. 

We ran the model both with and without the 17 secondary galaxy candidates in our sample. Both sets of parameters are listed in Table \ref{tab:allparams} for comparison. Removing the sub-sample of galaxies without velocity dispersion-based confirmation of dark matter had a negligible effect on the model parameters. The galaxy candidates are mostly very faint and populate the lower end of the distribution of galaxies without clusters, and so their removal does not significantly impact the logistic portion of the model. None of the candidates have cluster populations, and so they also do not influence the linear portion. We choose to leave the candidates in our sample for the remainder of our analysis. 

\subsection{Luminosity model results} \label{subsec:lum_results}

\vspace{-\baselineskip}
\vspace{-\baselineskip}
\begin{deluxetable*}{cccccc}[t] \label{tab:allparams}
    \tablecaption{Global parameter values for model runs.}
    \tablehead{
    \colhead{Model Version} & \colhead{$\beta_0$} &  \colhead{$\beta_1$} & \colhead{$\gamma_0$} & \colhead{$\gamma_1$} & \colhead{$\sigma_{M/L}$} \\
    \colhead{\textit{Data Used}} & \colhead{(cred. int.)} & \colhead{(cred. int.)} & \colhead{(cred. int.)} & \colhead{(cred. int.)} & \colhead{(cred. int.)} 
    }

    \startdata 
         Luminosity & $\-11.17$ & 1.60 & $-1.91$ & 0.96 & -- \\
         \textit{all galaxies} & $(-15.57,-7.39)$ & $(0.66,1.57)$ & $(-2.45,-1.36)$ & $(0.90, 1.02)$ &\\
         \\
         Luminosity & $-10.95$ & 1.57 & $-1.91$ &$0.96$ & -- \\
         \textit{confirmed galaxies} & $(-15.54,-6.92)$ & $(0.96,2.26)$ & $(-2.46,-1.35)$ & $(0.90,1.02)$ &\\
         \\
         HERBAL & $-11.05$ & 1.58 & $-0.88$ &  $0.85$ & $0.59$ \\
         \textit{all galaxies} & $(-15.55,-7.26)$ & $(0.99,2.27)$ & $(-2.21,0.38)$ & $(0.68,1.01)$ & $(0.38,0.89)$ \\
         \\
         HERBAL & $-14.93$ & 2.18 & $-0.58$ & $0.81$ & $0.60$ \\
        \textit{excluding outliers} & $(-20.89,-9.67)$ & $(1.36,3.10)$ & $(-2.17,0.94)$ & $(0.62,1.01)$ & $(0.38,0.92)$
    \enddata
    \tablecomments{Values cited are means and 95\% credible intervals.}
    \vspace{-3mm}
\end{deluxetable*}

\begin{figure*}
    \centering
    \includegraphics[width=0.99\textwidth]{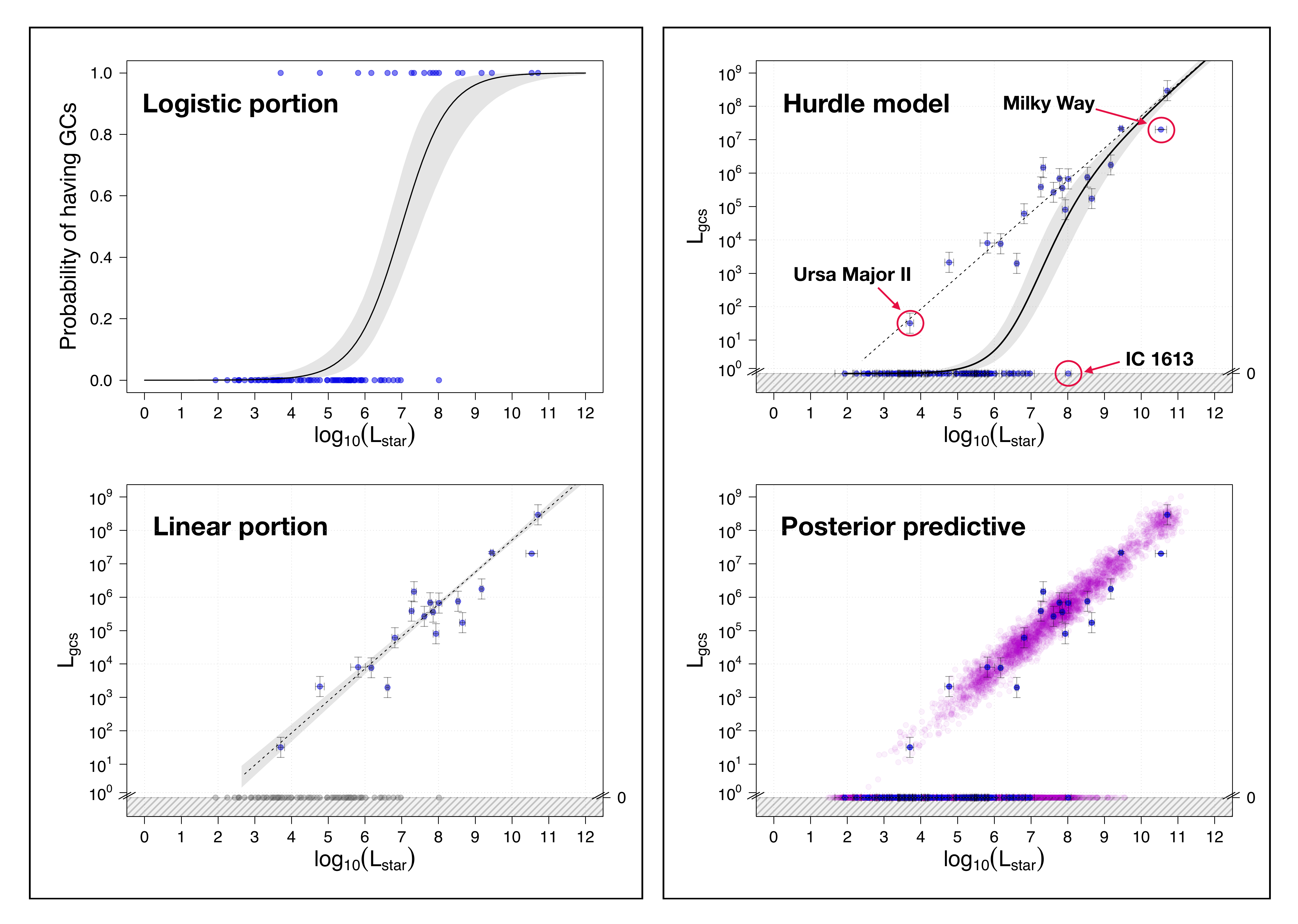}
    \caption{Results from the errors-in-variables hurdle model run on the full sample of Local Group galaxies. In all plots, 90\% credible intervals are colored in grey. In plots that show $L_{GCS}$ on a $log_{10}$ scale, zero values are added with an axis break for visualization purposes. \textbf{[Upper left]} The logistic portion of the luminosity hurdle model. The two logistic parameters given in Table \ref{tab:allparams}, which create this curve, give the probability that a galaxy with a given luminosity has GCs. \textbf{[Lower left]} The linear portion of the model. The linear parameters give the expected total luminosity of a galaxy's GC population, given that it has GCs (i.e. only the non-zero values are fitted). Zero values are greyed out. \textbf{[Upper right]} The expectation value of the Bayesian lognormal hurdle model. The hurdle model is a combination of the logistic and linear portions of the model shown on the left, and gives the overall expectation value for the mass of a galaxy's cluster system. The linear portion of the model is reproduced on this plot for comparison, and three interesting galaxies are labelled for further discussion. \textbf{[Lower right]} Posterior predictive plot containing data simulated from the model in purple with the Local Group sample plotted in blue. The two separate populations of galaxies - those with and without GCs - that are clearly seen in the simulated data.}
    \label{fig:lum}
\end{figure*}

The population level parameter estimates are reported in the first row of Table \ref{tab:allparams}. The mean values for each of the model parameters are reported along with 95 percent credible intervals.

The model fit is shown in Figure \ref{fig:lum}. The model can be interpreted as a two-step process to analyze the fit: the logistic portion of the model gives the probability of whether a galaxy has a cluster population or not, and the linear portion gives the expected value of the luminosity of the GC system, \textit{given} that the galaxy does have a non-zero number of GCs. These two portions of the model, plotted separately, are shown on the left hand side of Figure \ref{fig:lum}. The upper plot is the logistic portion of the model, which is probabilistic, and the lower plot is the linear portion of the model, which is a fit to the non-zero values (although it is plotted on $\log_{10}$ axes, the zero values are included with an axis break for comparison, but are greyed out). The grey shaded regions indicate the 90\% credible intervals for each portion of the fit. 

The logistic and linear models combine to make the hurdle model. Its expectation value $E[\log{L_{gcs}}]$ is shown in the upper right of Figure \ref{fig:lum} as a solid black line along with its 90\% credible interval shaded in grey. The median model approaches the linear slope (dotted line) at high masses where the probability of having clusters is almost 1, and closely follows the logistic curve at smaller masses where the probability of having clusters approaches zero. We stress that the solid black line is the \textit{expected value} of the model, and incorporates the probability that a galaxy has clusters. It does not represent values for any individual galaxy; while the model is bimodal, the expectation value is continuous and thus averages the expectation values of the two modes. 

In the lower right of Figure \ref{fig:lum} we show a posterior predictive plot, which is helpful in assessing model fit. We simulate 40,000 galaxies from our hurdle model and plot them in purple. Unlike the expectation value plot above, the posterior predictive clearly shows the bimodality of the model. There is a population of galaxies without GCs as well as a population with GCs that is linear as a function of luminosity. 

To simulate the data, we first randomly select chains from the MCMC and use the parameter values of the chains, $\theta_j$, to simulate data points. For each $\theta_j$, we randomly select $\log{L_{*,k}}$ values from the empirical cumulative distribution function (ECDF) of the Local Group galaxy luminosities. We calculate the probability $p_j$ that each galaxy $\log{L_{*,k}}$ has clusters using an inverse logit function with parameters $\beta_{0,j}, \beta_{1,j}$ from the chain $\theta_j$, and then randomly choose whether they do or do not have clusters with a Bernoulli distribution $\mathrm{Bern}(p_j)$. For the galaxies that do have clusters, we find their mass using the linear parameters $\gamma_{0,j}, \gamma_{1,j}$. We then inject scatter in both $\log{L_*}$ and $\log{L_{GC}}$ into the simulated data based on average observational uncertainties.

The simulated data in the posterior predictive plot appears underdispersed compared to the Local Group sample; many of the galaxies lie on the edges or outside the distribution of simulated galaxies. To quantitatively compare the dispersion of the linear portion of our model to our data, we calculate the standard deviation of the $\log{L_{gcs}}$ residuals of the non-zero populations. The standard deviation of the residuals of the Local Group galaxies (that have GCs) is 0.600, while the standard deviation of the posterior predictive is 0.291, or about half of that of the data.

The only uncertainty included in this model is observational uncertainty on galaxy and GC luminosities. The underdispersion of the simulated data indicates that our assumptions may not be accurate: we could be underestimating or mismodelling the observational uncertainties (e.g. our normal distribution assumption may be incorrect), or the $\log{L_*}-\log{L_{gcs}}$ relationship may contain either intrinsic scatter due to physical and evolutionary differences or other sources of uncertainty. In Section \ref{sec:lumtomass}, we will develop the HERBAL model, a hierarchical mass model that allows for further types of uncertainty.

\section{A Hierarchical Mass Model} \label{sec:lumtomass}

\begin{figure*} 
    \centering
    \includegraphics[width=0.95\textwidth]{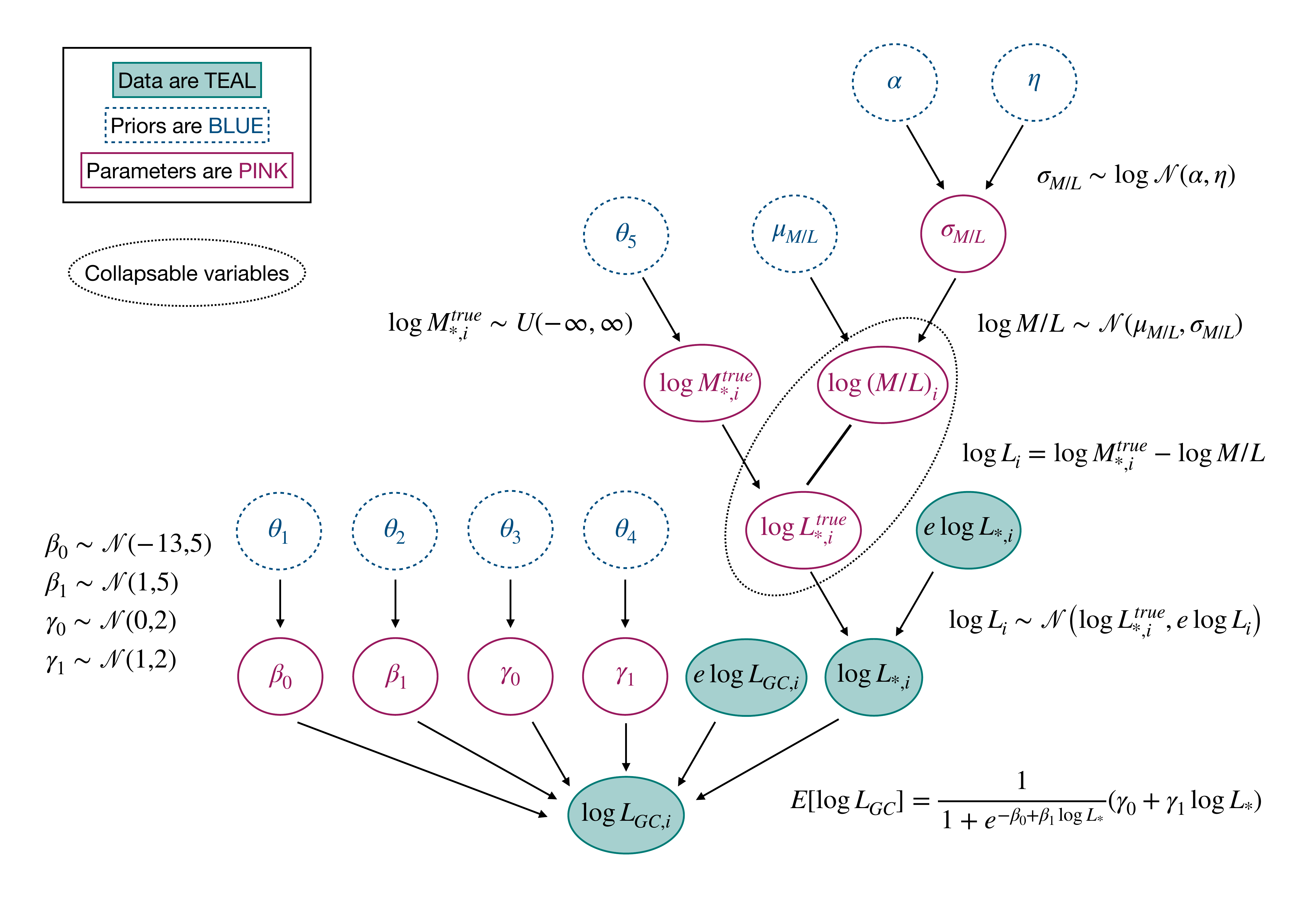}
    \caption{A directed acyclic graph (DAG) of the hierarchical model. The model predicts globular cluster system luminosities (the response variable) with a set of model parameters and galaxy luminosities (the predictor variable). The measured galaxy luminosity data are modeled by a normal distribution that depends on the true luminosities and the measurement error. The true luminosities are random variables which are calculated with true stellar masses and a M/L ratio, both of which are sampled for in the model. The M/L ratios for each galaxy are sampled from a global normal distribution of V-band M/L ratios which has a prior-set mean and a standard deviation that is fit by the model. The model can be collapsed along the true luminosities and M/L ratios since these are nuisance parameters, the true luminosity values are deterministic, and the normal distributions can be convolved. See Figure \ref{fig:dag_mass_collapsed} for the collapsed version of the model.}
    \label{fig:dag_mass}
    \vspace{3mm}
\end{figure*}

Although masses cannot be directly measured as luminosities are, they are a more physically relevant quantity in terms of galaxy evolution. Previous studies \citep[e.g.][]{Hudson2014, Harris2017, Choksi2019, Bastian2020} have investigated the relationship between (1) halo mass and globular cluster system mass, which is remarkably linear and gives constraints on cosmological models of galaxy formation and growth, and (2) stellar mass and globular cluster system mass, which is less linear but more easily inferred through galaxy luminosity, without requiring kinematic information or assumptions from cosmological models. To allow for comparisons between this analysis and previous works while avoiding additional sources of uncertainty, we now model the stellar mass relation instead of the luminosity relation or the halo mass relation.

Additionally, the errors-in-variables luminosity hurdle model created in Section \ref{sec:hurdle} is under-dispersed compared to the data. Another goal of the hierarchical mass model we develop here is to account for unmeasured scatter in the galaxy-GC relationship through the addition of more model parameters.

\citetalias{Eadie2022} models stellar masses from luminosity data using a color-corrected $M/L$ ratio \citep{Bell2003}, and then runs the model on the calculated masses. To account for uncertainties in the mass to light conversion, we instead add levels to our hierarchical luminosity model that assume a global distribution of V-band $M/L$ ratios and sample for the $M/L$ ratio and stellar mass of each galaxy in the sample. This method accounts for uncertainties not only in the luminosity measurements, but also in the $M/L$ ratios used for the galaxies. 

The HERBAL model is only hierarchical on the predictor variable (galaxy stellar masses). Instead of sampling for M/L ratios and stellar mass values for globular cluster systems as we do for the galaxies, we simply use a standard M/L ratio for all GC systems as described in Section \ref{subsec:sample_clusters}. We use the mass estimates and uncertainties reported for the GC systems of galaxies that use complex methods to estimate system mass, and the standard $M/L$ ratio of 1.88 for all other galaxies, giving them an uncertainty of a factor of two. 

\begin{figure}
    \centering
    \includegraphics[width=0.45\textwidth]{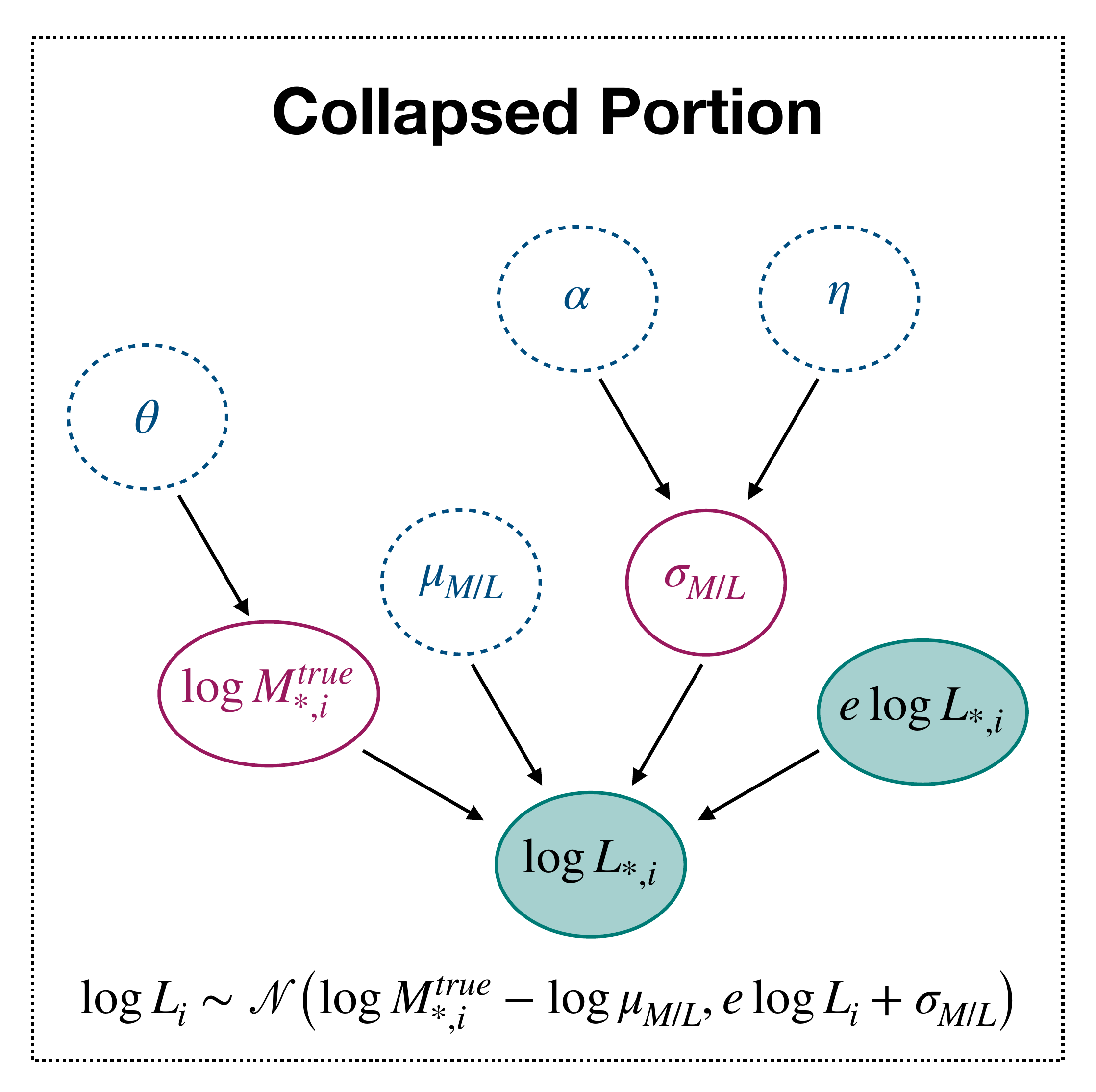}
    \caption{The collapsed portion of the model. The true luminosities and M/L ratios, which are nuisance parameters, have been removed from the model. This shifts the rest of the model downwards and combines the distributions from these levels of the model, which is feasible because the distributions describing these nuisance parameters are all either deterministic or normal distributions. This simplified, collapsed version of the model is the version actually implemented.}
    \label{fig:dag_mass_collapsed}
\end{figure}

\subsection{Structure of the hierarchical model}

The HERBAL model is shown in Figure \ref{fig:dag_mass}. This DAG has the same format as the one for the luminosity model (Figure \ref{fig:dag_lum}). The bottom three levels of the HERBAL model are identical to the luminosity model. However, it is hierarchical in galaxy luminosities, and so further levels are introduced which sample for galaxy masses and mass-to-light ratios for each galaxy. 

The true masses $\log{M_{*,i}^{true}}$ are sampled for in the model and have some prior $\theta$. The mass to light ratios for each galaxy, $\log{(M/L)_i}$, are sampled by the model and come from a normal distribution of global galactic $M/L$ ratios. We add a further level of hyperparameters by sampling for a global value of $\sigma_{M/L}$, the standard deviation of the distribution of $M/L$ ratios. Without further data, for example star formation histories of individual galaxies, the mean of the M/L distribution $\mu_{M/L}$ is entirely prior dominated, so we treat it as a fixed value. 

The $M/L$ ratios, mass values, and true luminosity values for each galaxy are connected via the $M/L$ ratio equation, making this level of the model deterministic. Unlike other levels of the model, where the levels above a quantity describe the distribution of that quantity, here the masses and $M/L$ ratios exactly determine the value of the true galaxy luminosity. 

The HERBAL model can be simplified for implementation by taking advantage of the fact that it contains multiple levels of Gaussians and a deterministic equation. The two circled nodes in the graph can be collapsed, which in essence removes them from the graph and shifts the upper levels of the model down based on the corresponding arrows. The collapsed subsection of the model is shown in Figure \ref{fig:dag_mass_collapsed}. In doing this, the Gaussians at each collapsed level are convolved. This leads to the distribution for $\log{L_*,i}$ shown in Figure \ref{fig:dag_mass_collapsed}, which contains all four input values by adding the means and standard deviations of the individual normal distributions in the hierarchical version. 

\subsection{Prior sensitivity and fixed values}

We leave the priors on the hurdle model parameters $\beta_0,\beta_1,\gamma_0,\gamma_1$ the same as those used in the luminosity model, described in Section \ref{subsec:lum_priors}. We have no further information about the parameter values that would lead us to tighten these priors. 

The V-band M/L distribution was chosen to be normal, with a mean $\mu_{M/L}$ and standard deviation $\sigma_{M/L}$. We fix $\mu_{M/L}$ at a value of 1. \citetalias{Eadie2022} calculated M/L ratios for the galaxies in the Local Group using a color-based equation from \citet{Bell2003}, and most of the galaxies had M/L ratios around 1 or slightly above. 

We make the scatter parameter $\sigma_{M/L}$ a model parameter so that the data can influence its value. Since $\sigma_{M/L}$ is a standard deviation, it must be positive. Galaxies typically have M/L ratios on the order of $1-10$, although some outliers that are dominated by low mass stars can have values an order of magnitude higher. We set a lognormal prior of $\sigma_{M/L}\sim \log{\mathcal{N}}(\alpha=0,\eta=0.5)$. The choice of a lognormal prior prevents negative parameter values, making 0 a lower limit on the value of $\sigma_{M/L}$. We tested a variety of other priors, both different lognormal distributions and uniform distributions, and found the best-fit values of $\sigma_{M/L}$ to be robust to these changes.

The true galaxy masses $\log{M_{*,i}^{true}}$ can also be assigned a prior, denoted as $\theta$ in Figure \ref{fig:dag_mass}. We choose to make $\theta$ uniform and unconstrained in this iteration of the model. A prior on galaxy masses could include information on galaxy type, total mass (often estimated for local group objects with kinematic tracers), or star formation rate, but due to the complex nature of any prior of this sort, we leave a more rigorous determination of $\theta$ to future work. 

\begin{figure*} 
    \centering
    \includegraphics[width=0.99\textwidth]{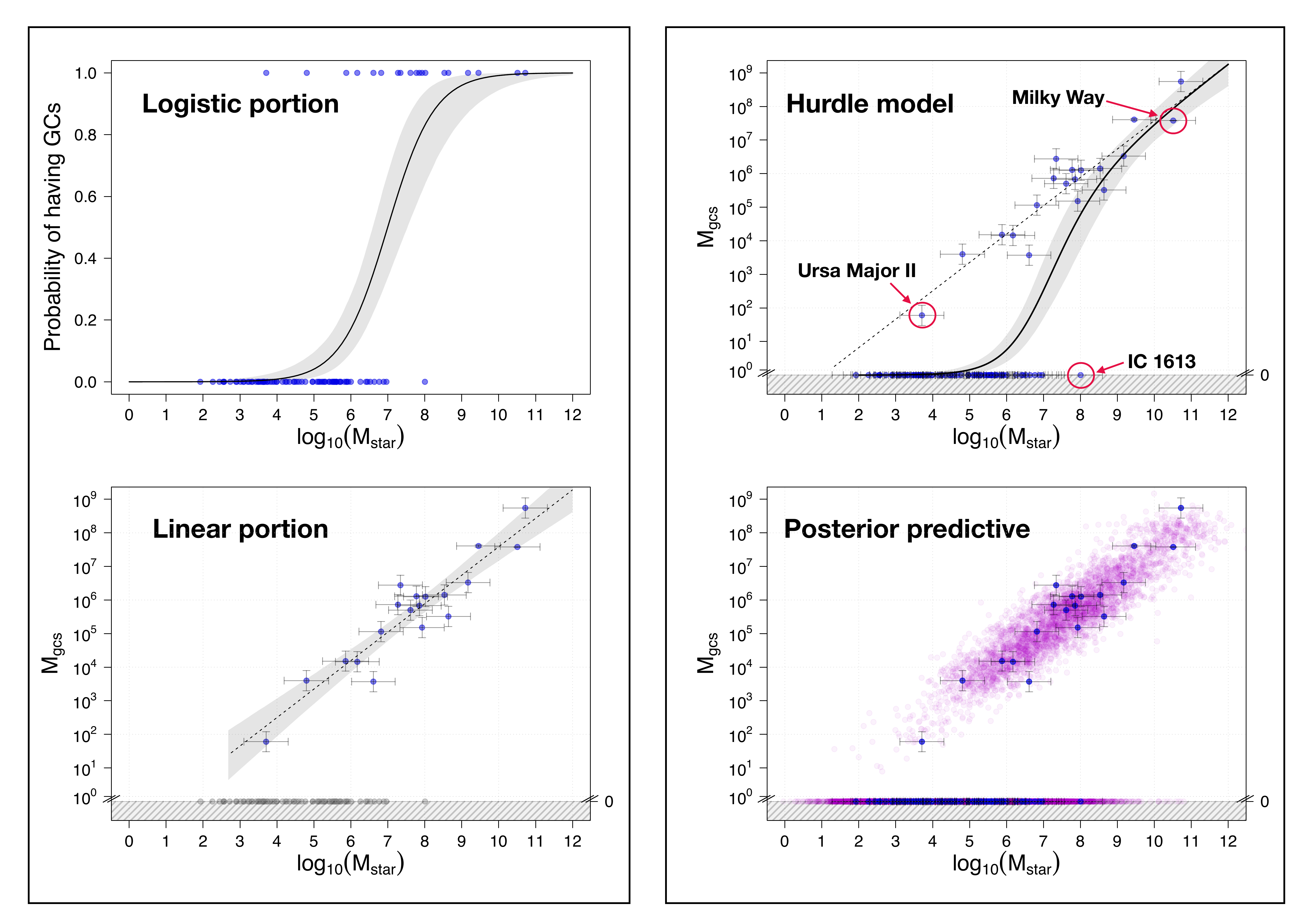}
    \caption{Results from the hierarchical hurdle model run on the full sample of Local Group galaxies. The stellar masses plotted are the best fit values sampled by the model, and the GC system masses are converted from luminosities using a M/L ratio of 1.88. Grey shaded regions indicate 90\% credible intervals. In plots that show $M_{GCS}$ on a $log_{10}$ scale, zero values are added with an axis break for visualization purposes. \textbf{[Upper left]} The logistic portion of the hurdle model, which models the probability that a galaxy does or does not have clusters. \textbf{[Lower left]} The linear portion of the hurdle model, which shows the expected size of a non-zero cluster population. Zero values are greyed out. \textbf{[Upper right]} The expectation value of the entire hurdle model with the linear fit over-plotted as a dotted line for comparison. Galaxies of interest are labelled and discussed individually in Section \ref{sec:disc}. \textbf{[Lower right]} Posterior predictive plot with data simulated from the hurdle model fit in purple.}
    \label{fig:mass_model_results}
    \vspace{3mm}
\end{figure*}

\subsection{Results} \label{subsec:mass_results}

The best-fit parameter values are reported in the third row of Table \ref{tab:allparams}. Both of the logistic $\beta$ parameters are very similar to those of the luminosity model. The $\gamma$ best-fit parameter values differ between the two models at the one sigma level. However, in both cases the solutions show that, \emph{for the galaxies that have GCs}, the total  GC luminosities (or masses) are nearly directly proportional to host galaxy luminosity (or mass) in log space.  For galaxies more massive than the Milky Way, the dependence takes on a different slope \citep{Harris2013}.

Figure \ref{fig:mass_model_results} shows the resulting model fit from the HERBAL model run on the entire Local Group sample. The galaxy stellar masses plotted are the mean values sampled from the chains. The GC masses are directly converted from luminosities using the standard $M/L$ ratio of $1.88$ as discussed above. Uncertainties on the galaxy stellar masses are a combination of observational uncertainties on the luminosity data and the model parameter $\sigma_{M/L}$, which encompasses uncertainty in the $M/L$ ratio of each galaxy as well as any other undefined sources of uncertainty or scatter in the data. Grey regions on each of the fits show the 90\% credible interval. Plots with a $\log_{10}(M_{GCS})$ axis have an axis break to include the zero values for visualization purposes.

The upper and lower left subplots show the logistic and linear parts of the fit, respectively. The logistic fit has a wide transition region, with substantial overlap in galaxy stellar mass between the population of galaxies that do not have GCs and those that do. The linear model also seems to fit the data well, especially with the expanded error bars that this model provides.

The upper right plot shows the expectation value of the mass hurdle model. Once again, we emphasize that the expectation value is a population-level value that incorporates the probability that a galaxy with a given stellar mass will have GCs as well as the expected mass of its GC population. It does not represent the expected size of the GC population for any one galaxy.

We again create posterior predictive samples to assess the model fit in the lower right hand plot. We simulate data from the mass model as we did from the luminosity model, with a few alterations. We randomly select $M_*$ values from the ECDF of the sampled galaxy masses for each randomly chosen chain $\theta_j$ instead of from the distribution of our data. We again randomly choose whether each galaxy has clusters based on a calculated probability of having clusters at the given stellar mass. We calculate the mass of those galaxies which do have clusters using the linear portion of the model, and once again add scatter to our points. However, the scatter added to this simulated data is based on the total combined uncertainty - both observational uncertainties and also the sampled values of $\sigma_{M/L,j}$.

The simulated data is shown along with the Local Group data in the lower right of Figure \ref{fig:mass_model_results}. 40,000 galaxies are simulated, and one sigma error bars are plotted on the data. The addition of a secondary source of uncertainty scatters the simulated data from the median relation far more than is seen in the luminosity model (Figure \ref{fig:lum}). Comparing the spread of residuals as we did for the luminosity model, we find that the standard deviation of the residuals of the Local Group data is 0.565, while the standard deviation of the posterior predictive residuals is 0.627. Unlike the luminosity model, this hierarchical mass model has a similar dispersion to the data, indicating that it is well dispersed.

\vspace{\baselineskip}
\section{Discussion} \label{sec:disc}

\subsection{Intrinsic scatter in the galaxy-GC relationship}

In the luminosity model, the only sources of uncertainty are the measurement uncertainties in both galaxy and globular cluster luminosities. The HERBAL model, on the other hand, samples for an additional global parameter, $\sigma_{M/L}$, which represents the scatter in mass to light ratios. Although this is its primary purpose, it is simply a parameter for otherwise unmeasured scatter, and thus has the ability to lump in other sources of uncertainty, such as any intrinsic scatter. Scatter in the (halo) mass - GC system mass relationship has been quantified in past studies \citep[e.g.][]{Harris2013, Choksi2019}, using using measures of scatter from linear regressions, but not with a Bayesian framework. 

Comparing the simulated data in the posterior predictive plot from the HERBAL model in Figure \ref{fig:mass_model_results} with the simulated data from the luminosity model in Figure \ref{fig:lum} shows the necessity of accounting for extra sources of uncertainty in the $M_*-M_{gcs}$ relation. The luminosity posterior predictive, lacking any method for accounting for uncertainty or scatter other than measurement error, is under-dispersed. The standard deviation of the residuals of the galaxies in the posterior predictive is only about half that of the Local Group galaxies, indicating that the Local Group galaxies are more scattered about the linear portion of the hurdle relationship than the model accounts for. The simulations from the HERBAL model, which allows for other, unmeasured uncertainty, are well dispersed, with a similar standard deviation of residuals to that of the data. We note that we have assumed normal error distributions for both the measurement uncertainties and the unmeasured error, which may be a faulty assumption. The $\sigma_{M/L}$ parameter, along with accounting for multiple sources of uncertainty in the model, may be compensating for these assumptions. Future work could investigate whether other error distributions could account for some of the underdispersion of the luminosity model.

\begin{figure}[t]
    \centering
    \includegraphics[width=0.45\textwidth]{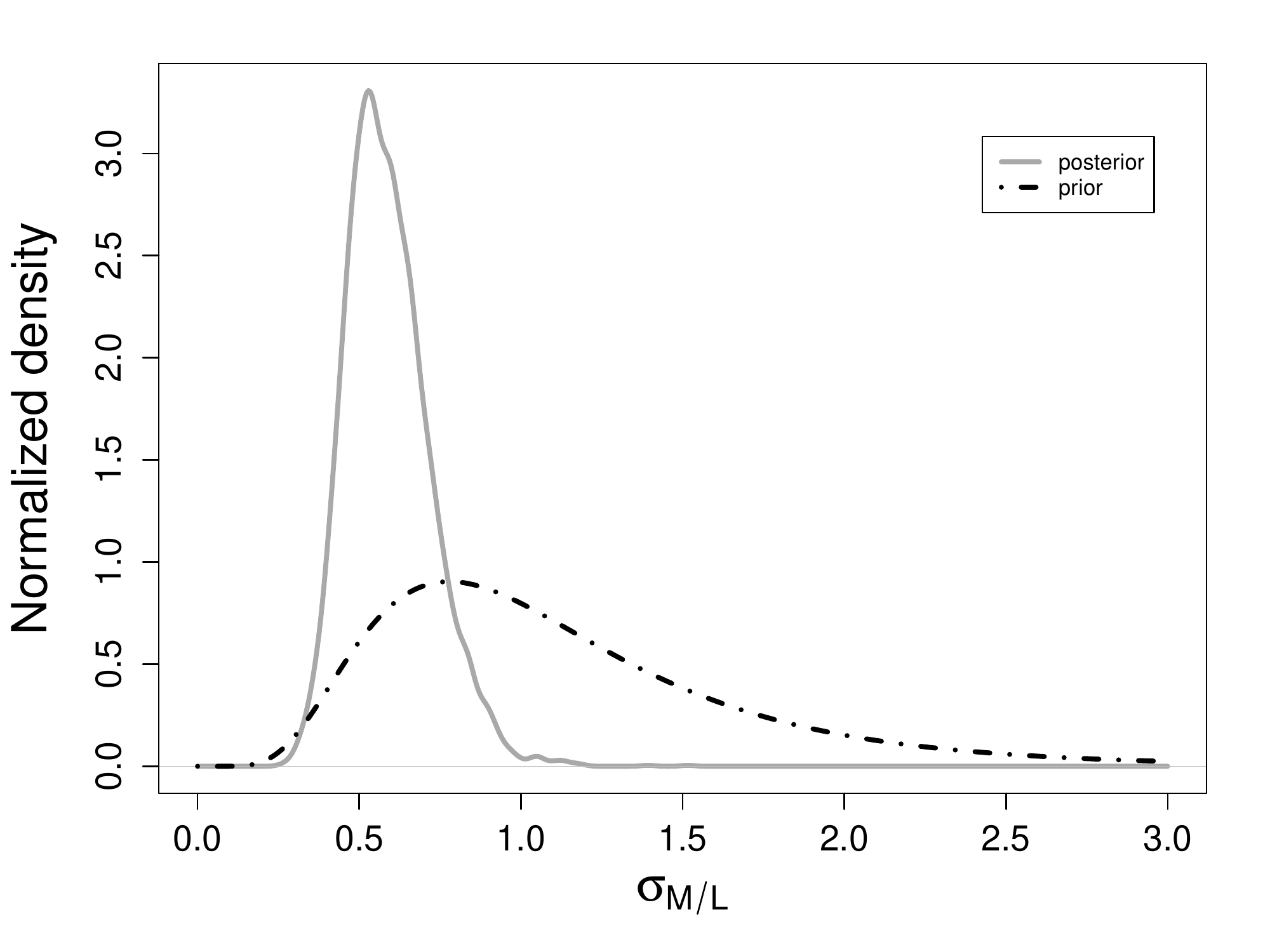}
    \caption{Posterior plot of the global model parameter $\sigma_{M/L}$. The lognormal prior is plotted as a black dashed line, while the posterior is solid grey. The posterior is constrained between $\sigma_{M/L}$ values of approximately 0.3 and 1, with a mean of 0.59.}
    \label{fig:sigma_posterior}
\end{figure}

The posterior of $\sigma_{M/L}$ is plotted in Figure \ref{fig:sigma_posterior}, along with its lognormal prior. The posterior (grey solid line) is more tightly constrained with a lower peak value than the prior (black dashed line). This indicates that the likelihood of the data showed less scatter about the mean M/L relationship than we assumed in the prior. The posterior has a strong peak between $\sigma_{M/L}$ values of 0.3 and 1.0, with almost no probability outside this range. Importantly, the posterior shows that $\sigma_{M/L}$ is not any smaller than about 0.3, and does not have a value near zero. A zero value would indicate that, under the assumed error model, all of the scatter in the $M_*-M_{gcs}$ relationship is due to measurement uncertainties on luminosities, and so this shows that there is measurable scatter in $M/L$ ratios and/or intrinsically in the relationship.

The posterior plot of $\sigma_{M/L}$, along with a comparison of the posterior predictive plots for both models, shows the importance of intrinsic scatter in the galaxy mass - GC system mass relation. Models of this relation should include a scatter term to account for this intrinsic feature of the relationship. Simulating data from the model fit gives a simple, visual method for checking dispersion and thus evaluating the model.

\subsection{M/L ratios in the hierarchical model}
The hierarchical model framework estimates values not just for global parameters like the hurdle model parameters $\beta_0, \beta_1, \gamma_0, \gamma_1$ and the global scatter $\sigma_{M/L}$, but also for parameters of individual galaxies such as galaxy stellar masses and M/L ratios. Since we have collapsed the model for easier implementation, we do not sample for nuisance parameters such as M/L ratios for individual galaxies, but an un-collapsed model could also sample values for these parameters. However, we do sample for stellar masses, and can therefore calculate approximate $M/L$ ratios from these and the luminosity data. 

\begin{figure}
    \centering
    \includegraphics[width=0.45\textwidth]{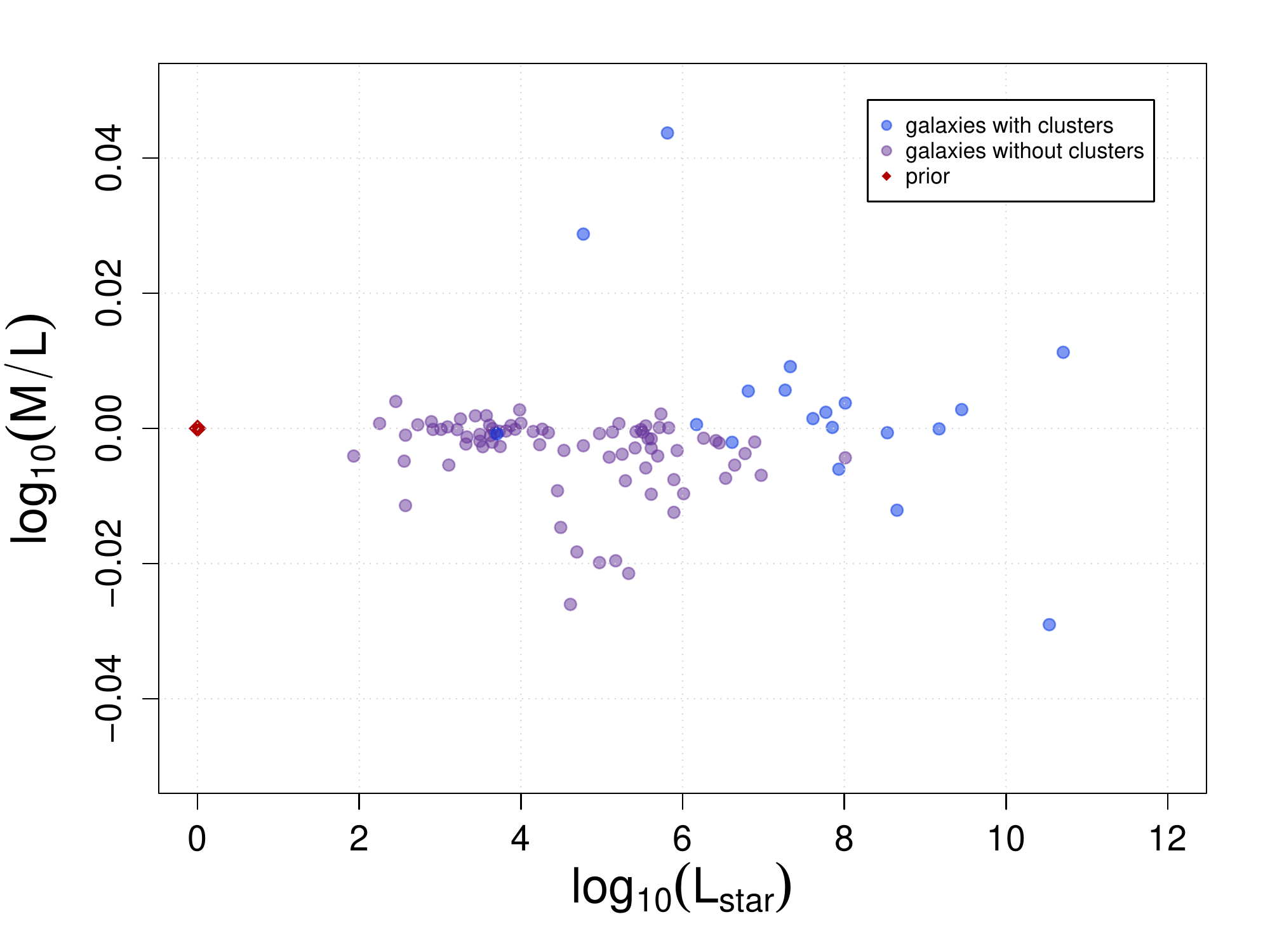}
    \caption{Mean mass-to-light ratios sampled by the hierarchical model for the Local Group galaxies. All galaxies have M/L ratios of approximately 1 ($\log{M/L}\approx 0$), which is near what is expected based on \citet{Bell2003}.}
    \label{fig:ML_ratios}
\end{figure}

Mass to light ratios for each galaxy are plotted against the galaxy luminosities in Figure \ref{fig:ML_ratios}. We chose 1 to be the fixed value for the mean $\mu$ of the distribution of $M/L$ ratios, based on calculations in \citetalias{Eadie2022} and expected values for $V$ band data and $(B-V)$ colors from \citet{Bell2003}. Figure \ref{fig:ML_ratios} shows that all galaxies in our sample have M/L ratios of approximately 1 (0 in log space). These values are in agreement with predictions from \citet{Bell2003}, indicating that our model is working well in this regard.

In general, galaxies without globular clusters have $M/L$ ratios that trend slightly below 0 in the log plot, while galaxies with clusters are spread more evenly around zero. This acts to slightly decrease the masses of galaxies without clusters compared to their luminosity values. Since less massive galaxies have a higher probability of being a zero, the model is pulling galaxies without GC populations down in mass to a higher probability case. Conversely, the smallest galaxies that do have clusters have the highest $M/L$ ratios, indicating that the model is pushing their masses higher because more massive galaxies are more likely to be non-zero. 

\subsection{Outliers in the sample: Ursa Major II and IC 1613}

\begin{figure*}
    \centering
    \includegraphics[width=0.99\textwidth]{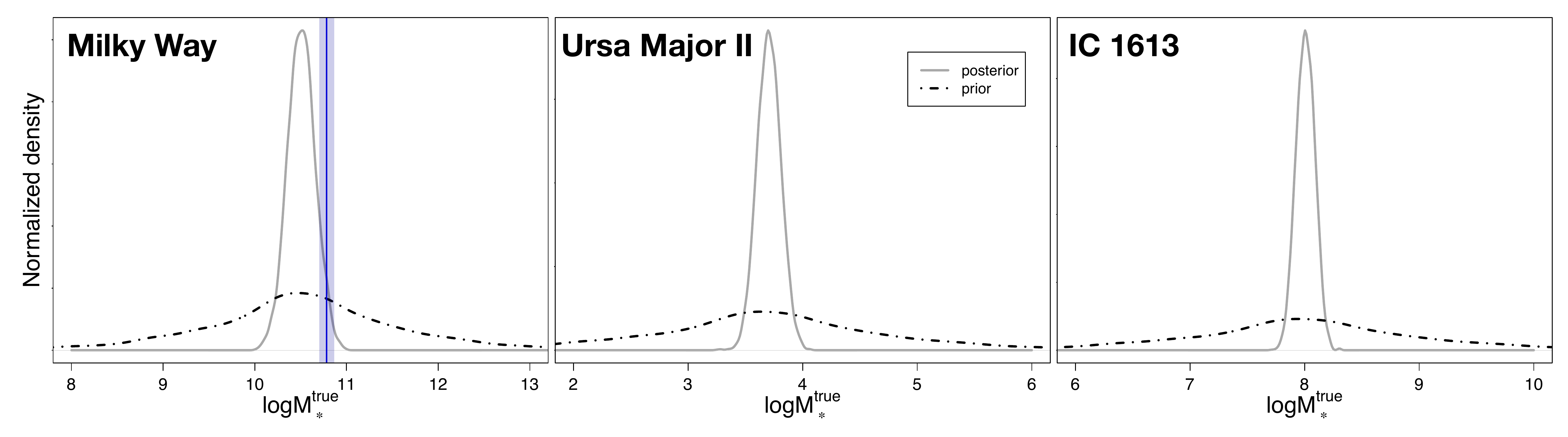}
    \caption{Posterior plots for the masses of three of the galaxies in our sample. We did not place a direct prior on the galaxy masses, so priors here are calculated from the prior on the $M/L$ distribution and $\sigma_{M/L}$. Ursa Major II and IC 1613 are the faintest galaxy to have GCs and the most luminous galaxy that doesn't have GCs, respectively. Their posteriors are both more constrained than the priors, but centered on the prior values. The Milky Way is of particular interest due to the amount it has been studied. A stellar mass estimate from \citet{Licquia2015b} is plotted along with our posterior and prior.}
    \label{fig:galaxy_posteriors}
\end{figure*}

There are two galaxies in our Local Group sample - IC 1613 and Ursa Major II - that appear to be either the most extreme galaxies in the normal population or are outliers. Both galaxies are labelled in Figures \ref{fig:lum} and \ref{fig:mass_model_results}. Here, we briefly discuss these galaxies individually. 

Ursa Major II is by far the least massive galaxy to host a globular cluster. Its luminosity is only $10^{3.704}L_\odot$, and its stellar mass is found to be $10^{3.703}M_\odot$. Its singular globular cluster is a central cluster that has a stellar mass (lower limit) of $52M_\odot$. The cluster is discussed in detail in \citetalias{Eadie2022}, and we refer interested readers to Appendix A of that paper for an in-depth discussion of the cluster. Since the cluster is a central cluster, it may not be best defined as a globular cluster. However, we have defined globular clusters purely based on age and not position, and so we include the cluster in our sample.

We show the posterior for the stellar mass estimate of Ursa Major II in the middle panel of Figure \ref{fig:galaxy_posteriors}. We did not place a direct prior on the stellar masses of galaxies, and so we calculate a prior based on the prior placed on $M/L$ ratios. We sample from the prior on $M/L$ ratios by first taking samples from the hyperprior $\sigma_{M/L} \sim \log{\mathcal{N}(0,0.5)}$ and then pulling samples from the prior on $M/L$ values: $M/L \sim \mathcal{N}(0, \sigma_{M/L})$. The stellar mass of Ursa Major II is well constrained by the model compared to the broad prior, and does not have a stellar mass (or $M/L$ ratio) that deviates significantly from that expected from the mean $M/L$. 

IC 1613 is the most massive galaxy in our sample to lack globular clusters. It is a dwarf irregular galaxy that is relatively isolated in the Local Group. It could plausibly be part of the M31 system, but also could be a fully isolated galaxy \citep{McConnachie2012}. Therefore, gravitational interactions with more massive galaxies are not a clear explanation for its lack of clusters. 

The posterior for the stellar mass of IC 1613 is shown in the right panel of Figure \ref{fig:galaxy_posteriors}. Similarly to Ursa Major II, the posterior is more constrained than the prior, with a mean mass estimate of $10^{8.008}M_\odot$. Its $M/L$ ratio is not an outlier from the rest of the sample (see Figure \ref{fig:ML_ratios}) and overall it behaves normally despite being the largest and most massive galaxy to have zero GCs.

Comparing IC 1613 with the posterior predictive reinforces the argument that it is not an outlier, and instead is simply the only galaxy in our sample to populate the high mass tail of the zero-cluster population of galaxies. The bottom right panel of Figure \ref{fig:mass_model_results} shows that the model predicts many galaxies of comparable and even larger masses to IC 1613 that do not have clusters. It appears as though it is our sample size, and not the intrinsic properties of IC 1613, that make it appear as an outlier. 

Removing these two galaxies from the sample changes the logistic parameters by about one sigma but does not noticeably change the linear parameters or $\sigma_{M/L}$. The global model parameters when the model is run without Ursa Major II and IC 1613 are listed in Table \ref{tab:allparams}. The difference in $\beta$ parameter values acts to steepen the logistic fit so that the transition region covers a smaller mass range. This is the expected effect from removing outliers.

\subsection{The Milky Way}

The Milky Way is an inherently interesting case study due to the vast amount of data we have on its internal structure and properties. Because of this, we look at the individual mass estimate of the Milky Way separately. The posterior for the Milky Way mass is shown on the left in Figure \ref{fig:galaxy_posteriors}. Its posterior looks similar to those of the other galaxies shown, and is more constrained than the prior. The blue vertical line is the stellar mass estimate (with uncertainty) from \citet{Licquia2015b}. Our posterior is not in disagreement with the Licquia value, but our mean value of $10^{10.507}M_\odot$ is smaller than the Licquia value of $10^{10.784}M_\odot$.

The Milky Way has one of the most uncertain luminosity measurements of our sample. Due to our position within the Milky Way, it is challenging to get a measurement of the integrated luminosity of the entire galaxy. The luminosity estimate used here, from \citet{Licquia2015}, is based on dependencies between stellar mass, star formation rate, and luminosity, using a sample of Milky Way analogs to compare to our galaxy. Without a direct measurement of galaxy luminosity, we acknowledge that our value could be biased, which could be affecting the mass estimated by the model.

The Milky Way is also unique in that we have strong independent constraints on $M_*$. The stellar mass of the Milky Way calculated by \citet{Licquia2015b} via a Baysian hierarchical model combining many previous mass estimates is larger than the mass found by our model. Our model is only dependent on the weak luminosity constraints we have for the Milky Way, without taking into account any of the independent, stronger mass constraints that exist. Therefore, the Galaxy is shifted toward the mean value of the linear portion of the hurdle model. In future work we aim to add the ability to incorporate multiple types of constraints (e.g. mass constraints) on different hierarchical steps of our model.

\section{Conclusions} \label{sec:conc}

In this work, we developed a hierarchical Bayesian lognormal hurdle model to describe the relationship between galaxies' stellar masses and their globular cluster system masses, fully accounting for sources of uncertainty. Our key findings are summarized below. 

\begin{enumerate}
    \item We compile a catalog of galaxies in the Local Group, their luminosities, and the number and combined luminosity of their globular cluster systems. We distinguish between confirmed galaxies and ultra-faint dwarf galaxy candidates, which do not have confirmation of an associated dark matter overdensity. The full catalog is presented in Appendix A and is publicly available. 
    \item We develop an errors-in-variables Bayesian lognormal hurdle model to incorporate measurement uncertainties in both galaxy and GC luminosities into the hurdle model first proposed in \citetalias{Eadie2022}. The best-fit coefficients of the logistic portion of the fit, which gives a probability that a galaxy at a given luminosity has clusters or not, are $(\beta_0=7.80, \beta_1=-1.09)$. The stellar mass at which 50\% of galaxies are expected to host non-zero globular cluster populations is $\log_{10}(M_*)=7.181$. The median coefficients for the linear portion of the model, which give the expected luminosity of a globular cluster system \textit{given} that the galaxy is known to have clusters, are $(\gamma_0=-1.91, \gamma_1=0.96)$. 
    \item To make predictions about globular cluster system mass based on galaxy stellar mass, we develop the HERBAL model, a hierarchical hurdle model. This model incorporates an additional uncertainty term as well as the measurement uncertainties used in the errors-in-variables model above. This unmeasured uncertainty term accounts for variation in M/L ratios and intrinsic scatter in the model. The hierarchical nature of the model also gives estimates for stellar masses for each galaxy, and by extension, estimates of M/L ratios. The parameters of this model are $(\beta_0=7.73, \beta_1=-1.07)$ and $(\gamma_0=-0.89, \gamma_1=0.85)$. The stellar mass at which 50\% of galaxies are expected to host globular cluster populations is $\log_{10}(M_*)=6.996$, which is similar to results from \citep{Chen2023}.
    \item We simulate data for posterior predictive plots for both the errors-in-variable model and the HERBAL model. The posterior predictive from the HERBAL model matches the Local Group data well, but the luminosity model is underdispersed. The only source of error in the luminosity model is measurement error, which is assumed to be normally distributed in log space, while the HERBAL model includes a parameter meant to model uncertainty in M/L ratios but that can also incorporate other sources of error in the model (also assumed to be normal). The underdispersion of the luminosity model indicates that our assumptions for the error distributions may be incorrect, and there could be intrinsic scatter in the galaxy-GC relation. 
    \item The linear portion of the hurdle model is robust to outliers, while the logistic portion is slightly affected by galaxies in the Local Group with high stellar masses and no globular clusters and those with low stellar masses that have clusters. There are two such galaxies in our sample: (1) IC 1613, which is an isolated, relatively massive irregular dwarf galaxy without GCs, and (2) Ursa Major II, which is a very small dwarf with a central star cluster. Removing both these galaxies shrinks the transition region of the hurdle model, steepening the logistic portion of the fit.
    \item Our estimate of the Milky Way's stellar mass is $10^{10.507}M_\odot$, which is lower than the \citet{Licquia2015b} estimate of $10^{10.784}M_\odot$. The Licquia and Newman estimate is, however, within our 95\% credible interval. Unlike many of the other galaxies in our sample, we have strong independent constraints on the Milky Way's stellar mass that are tighter than the constraints we have on its luminosity. This highlights a weakness of the current form of the HERBAL model, which requires luminosity data and does not allow for the incorporation of independent constraints on galaxy masses or M/L ratios.
\end{enumerate}

In this work, we demonstrate that a Bayesian lognormal hurdle model can describe the $M_*-M_{gcs}$ relation, especially for low-mass galaxies. The relationship has previously been modeled linearly, but this has led to discrepancies and uncertainty in the low mass regime. Our hurdle model (the HERBAL model) has the capability to simultaneously model the linear relationship between GC system mass and galaxy mass for galaxies which have clusters while also accounting for galaxies that do not have GCs. Our results indicate that the linear relationship appears to hold down to the lowest mass galaxies with clusters, while there is a separate population of low-mass galaxies without clusters. 

In future works, we will expand the HERBAL model to be more generalizable and take advantage of varied data types. The model can be expanded into an additive or mixed model which could separate between galaxy types or environments which may have different $M_*-M_{gcs}$ relations. A wider variety of inputs beyond just luminosity data can be incorporated to give the model more information about galaxies such as the Milky Way for which we have independent mass constraints. Other likelihoods that may better describe the variance in the data can be tested, as well as uncertainty distributions beyond normal and lognormal models. Testing a wider variety of models will help with understanding the extent to which intrinsic scatter is important in the $M_*-M_{gcs}$ relation. Each of these upgrades can be worked directly into the hierarchical model, just as the mass conversions and additional uncertainty were built upon the first errors-in-variables model. As the model is expanded, it will  give us a clearer view into the globular cluster populations of small galaxies, giving us hints as to how they formed and evolved. 

\section*{Acknowledgements}

SCB would like to thank Aaron Springford and Steffani Grondin for their insight in many useful conversations that have improved this work. She would also like to thank everyone whose questions and feedback during talks and presentations about this work led to the improvement of methods and visualizations.
WEH acknowledges the financial support of NSERC (Natural Sciences and Engineering Research Council of Canada). GE acknowledges funding from NSERC through Discovery Grant RGPIN-2020-04554.

\software{Stan Modeling Lanugage \citep{stan}, R Statistical Software Environment \citep{baser}, and the following R packages: \texttt{rstan} \citep{rstan}, \texttt{plotrix} \citep{plotrix}, \texttt{ggplot2} \citep{ggplot}, \texttt{latex2exp} \citep{rlatex}, \texttt{bayesplot} \citep{bayesplot}.}

\vspace{5mm}

\bibliography{paper}{}
\bibliographystyle{aasjournal}

\appendix

\section{Table of Local Group galaxies} \label{sec:app1}

\begin{longtable}{cccccccc}

\caption{Local Group Sample} \label{tab:sample} \\

\multicolumn{1}{c}{\textbf{Galaxy}} & \multicolumn{1}{c}{\textbf{Confirmed}} & \multicolumn{1}{c}{\textbf{$\log{L_*}$}} & \multicolumn{1}{c}{\textbf{$e\log{L_*}$}} & \multicolumn{1}{c}{\textbf{$N_{GC}$}} & \multicolumn{1}{c}{\textbf{$M_{*,GC}$}} & \multicolumn{1}{c}{\textbf{Galaxy Ref.}} & \multicolumn{1}{c}{\textbf{GC Ref.}} \\ \hline \hline
\endfirsthead

\multicolumn{8}{c}
{{\bfseries \tablename\ \thetable{} -- continued from previous page}} \\
\multicolumn{1}{c}{\textbf{Galaxy}} & \multicolumn{1}{c}{\textbf{Confirmed}} & \multicolumn{1}{c}{\textbf{$\log{L_*}$}} & \multicolumn{1}{c}{\textbf{$e\log{L_*}$}} & \multicolumn{1}{c}{\textbf{$N_{GC}$}} & \multicolumn{1}{c}{\textbf{$M_{*,GC}$}} & \multicolumn{1}{c}{\textbf{Galaxy Ref.}}  & \multicolumn{1}{c}{\textbf{GC Ref.}} \\ \hline \hline
\endhead

\hline \multicolumn{8}{r}{{Continued on next page}} \\
\endfoot

\endlastfoot

And I & 1 & 6.612 &  0.04 & 1 & $3.68 \times 10^3$ & 0 & 1 \\
Eridanus II & 1 & 4.772 &  0.12 & 1 & $4 \times 10^3$ & 1 & 1 \\
Aquarius & 1 & 6.172 & 0.04 & 1 & $1.46 \times 10^4$ & 0 & 1 \\
And XXV & 1 & 5.812 & 0.2 & 1 & $1.46 \times 10^4$ & 0 & 1 \\
Pegasus & 1 & 6.812 & 0.08 & 1 & $1.15 \times 10^5$ & 0 & 1 \\
SMC & 1 & 8.652 & 0.08 & 1 & $3.2 \times 10^5$ & 0 & 1 \\ 
WLM & 1 & 7.612 & 0.04 & 1 & $5.04 \times 10^5$ & 0 & 1 \\
IC 10 & 1 & 7.932 & 0.08 & 1 & $1.514 \times 10^5$ & 0 & 2 \\ 
NGC 185 & 1 & 7.852 & 0.04 & 8 & $6.8 \times 10^5$ & 0 & 1 \\
Fornax & 1 & 7.268 & 0.056 & 5 & $7.3 \times 10^5$ & 1 & 1 \\
NGC 6822 & 1 & 8.012 & 0.08 & 8 & $1.26 \times 10^6$ & 0 & 1 \\
NGC 147 & 1 & 7.772 & 0.04 & 10 & $1.3 \times 10^6$ & 0 & 1 \\
NGC 205 & 1 & 8.532 & 0.04 & 11 & $1.41 \times 10^6$ & 0 & 1 \\
Sagittarius & 1 & 7.332 & 0.06 & 8 & $2.78 \times 10^6$ & 1 & 1 \\ 
LMC & 1 & 9.172 & 0.04 & 16 & $3.3 \times 10^6$ & 0 & 1 \\ 
M33 & 1 & 9.452 & 0.04 & 81 & $4.058 \times 10^7$ & 0 & 3 \\
Milky Way & 1 & 10.536 & 0.156 & 161 & $3.79 \times 10^7$ & 2 & 4 \\
M31 & 1 & 10.708 & 0.1008 & 450 & $5.516 \times 10^8$ & 3 \footnote{http://leda.univ-lyon1.fr/ledacat.cgi?o=M31} & 5 \\ 
Ursa Minor & 1 & 5.544 & 0.02 & 0 & -- & 1 & -- \\
Virgo I & 0 & 2.252 & 0.36 & 0 & -- & 1, 4 & -- \\ 
Willman I & 1 & 3.092 & 0.296 & 0 & -- & 1 & -- \\
And VII & 1 & 6.972 & 0.12 & 0 & -- & 0 & -- \\
And X & 1 & 4.972 & 0.4 & 0 & -- & 0 & -- \\ 
And XI & 1 & 4.692 & 0.52 & 0 & -- & 0 & -- \\
And XII & 1 & 4.492 & 0.48 & 0 & -- & 0 & -- \\
And XIII & 1 & 4.612 & 0.52 & 0 & -- & 0 & -- \\
And XIV & 1 & 5.292 & 0.24 & 0 & -- & 0 & -- \\
And XIX & 1 & 5.612 & 0.24 & 0 & -- & 0 & -- \\
And XV & 1 & 5.692 & 0.16 & 0 & -- & 0 & -- \\
And XVI & 1 & 5.612 & 0.16 & 0 & -- & 0 & -- \\
And XVII & 1 & 5.412 & 0.16 & 0 & -- & 0 & -- \\
And XVIII & 1 & 5.612 & 0.16 & 0 & -- & 20 & -- \\ 
And XX & 1 & 4.452 & 0.44 & 0 & -- & 0 & -- \\ 
And XXI & 1 & 5.892 & 0.24 & 0 & -- & 0 & -- \\
And XXII & 1 & 4.532 & 0.32 & 0 & -- & 19 & -- \\ 
And XXIII & 1 & 6.012 & 0.2 & 0 & -- & 0 & -- \\ 
And XXIV & 1 & 4.972 & 0.2 & 0 & -- & 0 & -- \\
And XXIX & 1 & 5.252 & 0.16 & 0 & -- & 0 & -- \\
And XXVI & 1 & 4.772 & 0.2 & 0 & -- & 0 & -- \\
And XXVII & 1 & 5.092 & 0.2 & 0 & -- & 0 & -- \\ 
And XXVIII & 1 & 5.332 & 0.4 & 0 & -- & 0 & -- \\ 
Antlia II & 1 & 5.554 & 0.2 & 0 & -- & 5 & -- \\
Aquarius II & 1 & 3.676 & 0.056 & 0 & -- & 1 & -- \\
Bootes I & 1 & 4.34 & 0.1 & 0 & -- & 1 & -- \\
Bootes II & 1 & 3.108 & 0.3 & 0 & -- & 1 & -- \\ 
Bootes III & 0 & 4.232 & 0.2 & 0 & -- & 0, 6 & -- \\
Bootes IV & 0 & 3.744 & 0.2 & 0 & -- & 18 & -- \\
Carina & 1 & 5.712 & 0.02 & 0 & -- & 1 & -- \\
Carina II & 1 & 3.732 & 0.04 & 0 & -- & 1 & -- \\
Carina III & 0 & 2.892 & 0.08 & 0 & -- & 1, 7 & -- \\
Centaurus I & 0 & 4.154 & 0.044 & 0 & -- & 8 & -- \\
Cetus & 1 & 6.412 & 0.08 & 0 & -- & 0 & -- \\
Cetus II & 0 & 1.932 & 0.272 & 0 & -- & 1, 9 & -- \\
Cetus III & 0 & 2.912 & 0.228 & 0 & -- & 1, 4 & -- \\ 
Columba I & 0 & 3.612 & 0.08 & 0 & -- & 1, 10 & -- \\
Coma Berenices & 1 & 3.644 & 0.2 & 0 & -- & 1 & -- \\
Crater II & 1 & 5.212 & 0.04 & 0 & -- & 1 & -- \\
Canes Venatici I & 1 & 5.424 & 0.024 & 0 & -- & 1 & -- \\
Canes Venatici II & 1 & 4.00 & 0.128 & 0 & -- & 1 & -- \\ 
Draco & 1 & 5.484 & 0.02 & 0 & -- & 1 & -- \\
Grus I & 1 & 3.32 & 0.236 & 0 & -- & 1, 11 & -- \\
Grus II & 0 & 3.492 & 0.088 & 0 & -- & 1, 12 & -- \\
Hercules & 1 & 4.264 & 0.068 & 0 & -- & 1 & -- \\
Horologium I & 1 & 3.436 & 0.224 & 0 & -- & 1 & -- \\
And II & 1 & 6.892 & 0.08 & 0 & -- & 0 & -- \\
Horologium II & 0 & 2.556 & 0.408 & 0 & -- & 1 & -- \\
Hydra II & 0 & 3.876 & 0.148 & 0 & -- & 1, 13 & -- \\
Hyrdus I & 1 & 3.816 & 0.032 & 0 & -- & 1 & -- \\
IC 1613 & 1 & 8.012 & 0.08 & 0 & -- & 0 & -- \\
Leo A & 1 & 6.772 & 0.08 & 0 & -- & 0 & -- \\
Leo I & 1 & 6.644 & 0.112 & 0 & -- & 1 & -- \\
Leo II & 1 & 5.828 & 0.016 & 0 & -- & 1 & -- \\
Leo IV & 1 & 3.928 & 0.144 & 0 & -- & 1 & -- \\
And III & 1 & 5.932 & 0.12 & 0 & -- & 0 & -- \\
Leo V & 1 & 3.648 & 0.144 & 0 & -- & 1 & -- \\
Leo T & 1 & 5.132 & 0.2 & 0 & -- & 0 & -- \\ 
And IX & 1 & 5.172 & 0.44 & 0 & -- & 0 & -- \\
Pegasus III & 1 & 3.572 & 0.2 & 0 & -- & 1 & -- \\
Phoenix & 1 & 5.892 & 0.16 & 0 & -- & 0 & -- \\
Phoenix II & 0 & 3.012 & 0.16 & 0 & -- & 1, 14 & -- \\
Pictor II & 0 & 3.212 & 0.2 & 0 & -- & 1, 15 & -- \\
Pisces II & 1 & 3.624 & 0.152 & 0 & -- & 1 & -- \\
Reticulum II & 1 & 3.528 & 0.152 & 0 & -- & 1 & -- \\
Reticulum III & 0 & 3.252 & 0.116 & 0 & -- & 1 & -- \\
Sculptor & 1 & 6.26 & 0.056 & 0 & -- & 1 & -- \\
Segue I & 1 & 2.452 & 0.292 & 0 & -- & 1 & -- \\
And V & 1 & 5.572 & 0.08 & 0 & -- & 0 & -- \\
Segue II & 1 & 2.724 & 0.352 & 0 & -- & 1, 16 & -- \\
Sextans & 1 & 5.508 & 0.024 & 0 & -- & 1 & -- \\
Sagittarius dIrr & 1 & 6.532 & 0.12 & 0 & -- & 0 & -- \\
Triangulum II & 0 & 2.572 & 0.304 & 0 & -- & 1, 17 & -- \\
Tucana & 1 & 5.732 & 0.08 & 0 & -- & 0 & -- \\
Tucana II & 1 & 3.492 & 0.08 & 0 & -- & 1 & -- \\
Tucana IV & 0 & 3.332 & 0.112 & 0 & -- & 1, 12 & -- \\
And VI & 1 & 6.452 & 0.08 & 0 & -- & 0 & -- \\
Tucana V & 0 & 2.572 & 0.196 & 0 & -- & 12 & -- \\
Ursa Major I & 1 & 3.984 & 0.152 & 0 & -- & 1 & -- \\
Ursa Major II & 1 & 3.704 & 0.104 & 1 & 1.78 & 1 & 6 \\

\hline \hline 
\caption{\textbf{Column Descriptions:} (1) object name; (2) status, where confirmed galaxy = 1 and candidate galaxy = 0; (3) galactic log luminosity (calculated from V-band absolute magnitude); (4) uncertainty in the galactic log luminosity; (5) number of globular clusters; (6) total mass of the globular cluster system (7) references for the galaxy luminosity, where sources are as follows: 0. \cite{McConnachie2012}, 1. \cite{Simon2019}, 2. \cite{Licquia2015}, 3. \cite{Makarov2014}, 4. \cite{Homma2018}, 5. \cite{Torrealba2019}, 6. \cite{Carlin2018}, 7. \cite{Ji2020}, 8. \cite{Mau2020}, 9. \cite{Conn2018}, 10. \cite{Fritz2019}, 11. \cite{Chiti2022}, 12. \cite{Simon2020}, 13. \cite{Kirby2015}, 14. \cite{Mutlu-Pakdil2018}, 15. \cite{Drlica-Wagner2016}, 16. \cite{Kirby2013}, 17. \cite{Carlin2017}, 18. \cite{Homma2019}, 19. \cite{Martin2009}, 20. \cite{Martin2016}; (8) references for the globular cluster system luminosity or mass, where sources are as follows: 1. \cite{Forbes2018}, 2. \cite{Lim2015}, 3. \cite{Fan2014}, 4. \cite{Vasiliev2021}, 5. \cite{Galleti2004}, 6. \cite{Eadie2022}.}
\end{longtable}

\end{document}